\documentclass[%
 reprint,
 amsmath,amssymb,
 aps,
]{revtex4-2}

\usepackage{graphicx}
\usepackage{dcolumn}
\usepackage{bm}
\usepackage{amsmath}
\usepackage[usenames,dvipsnames]{xcolor} 
\usepackage{mathtools}
\usepackage[colorlinks=true,citecolor=blue,linkcolor=blue]{hyperref}
\usepackage[caption=false]{subfig}
\usepackage{bbm}
\usepackage{braket}
\usepackage{physics}
\usepackage{comment}
\usepackage{upgreek}
\usepackage{esint}
\usepackage{soul}
\usepackage[normalem]{ ulem } 

\newcommand{\leri}[1]{\left(#1\right)}

\begin{document}

\title{Fast design and scaling of multiqubit gates in large-scale trapped-ion quantum computers}

\author{Lee Peleg$^{1,3\ast}$}
\author{David Schwerdt$^{1,\ast}$}
\author{Jonathan Nemirovsky$^{1,3\ast}$}
\author{Yotam Shapira$^{1,3\ast}$}
\author{Nitzan Akerman$^{1}$}
\author{Ady Stern$^{2}$}
\author{Amit Ben Kish$^{3}$}
\author{Roee Ozeri$^{1}$}

\affiliation{\small{$^1$Department of Physics of Complex Systems, Weizmann Institute of Science, Rehovot 7610001, Israel\\
$^2$ Department of Physics of Condensed Matter Systems, Weizmann Institute of Science, Rehovot 7610001, Israel\\
$^3$ Quantum Art LTD, Ness Ziona 7403682, Israel\\
$^*$ These authors contributed equally to this work
}}

\begin{abstract}
    Quantum computers based on crystals of trapped ions are a prominent technology for quantum computation. A unique feature of trapped ions is their long-range Coulomb interactions, which can be exploited to realize large-scale multiqubit entanglement gates. However, scaling up the number of qubits, $N$, in these systems, while retaining high-fidelity and high-speed operations, is challenging. Specifically, designing multiqubit entanglement gates in long ion crystals of hundreds of ions involves an NP-hard optimization problem, rendering scale-up not only a technological challenge, but also a conceptual challenge. Here we introduce a method that mitigates this challenge, effectively allowing for a polynomial-time design of fast, robust, and programmable entanglement gates, acting on the entire ion-crystal. We show that while the number of simultaneous entanglement operations scales as $N^2$, the gate duration scales as $N$, leading to a scaling advantage. We use our methods to investigate the drive-power requirements and susceptibility to noise and errors of these multiqubit gates. Our method delineates a path towards scaling up quantum computers based on ion-crystals with hundreds of qubits.
\end{abstract}

\maketitle

\section{Introduction}
Trapped-ions are a leading quantum computation platform due to their accurate control, long-range connectivity, and long coherence times. Despite their all-to-all connectivity, linear ion crystals of growing lengths present increasing difficulty in implementing high-fidelity and high-speed entanglement gates. Some trapped ion scale-up architectures overcome this challenge by interconnecting separate ion crystals, either by ion shuttling between segments in a quantum charge-coupled device architecture \cite{kielpinski2002architecture,pino2021demonstration,moses2023race} or by photonic interconnects \cite{moehring2007entanglement,nadlinger2022experimental,krutyanskiy2023entanglement}. However, both approaches to scalability will benefit from working with longer ion crystals as their basic building block, taking full advantage of the inherent long-range connectivity of the ions and its expected benefits \cite{shapira2020theory,grzesiak2020efficient,schwerdt2022comparing,bravyi2022constant,bassler2023synthesis}.

Quantum information processing devices, based on crystals of 10 to 100s of trapped ions, have recently been proposed and implemented \cite{yao2022experimental,pogorelov2021compact,feng2022continuous,schwerdt2024scalable,guo2024site}, overcoming key hurdles such as crystal stability, cooling, and coherence. However, a prominent open challenge is the design and implementation of high-fidelity, programmable multiqubit entangling gates in these large-scale systems. In trapped-ion quantum computers, entanglement is mediated by the normal modes of motion of the ion crystal. As the number of ions increases, the motional spectrum becomes denser, which complicates the gate design. This challenge manifests itself as a quadratically constrained NP-hard optimization problem \cite{Blume1989Theory,shapira2020theory}, that scales as $N^2$ with the number of ions $N$ , making the implementation and the study of feasibility, performance, and scaling of such multiqubit gates a formidable challenge.

Crucially, solving, or approximately solving, this optimization problem enables the implementation of multiqubit entangling gates that offer programmable simultaneous interactions between all qubit pairs in the ion-crystal. Specifically, for a $N$ qubit register and given a desired set of $\binom{N}{2}=N(N-1)/2$ real-valued qubit-qubit couplings, $\varphi_{nn'}$, with $n,n'=1,...,N$, these gates couple to multiple modes of motion of the ion-crystal and generate the logical operation,
\begin{equation}
    U_\text{MQ}\leri{\varphi}=\exp\leri{i\sum_{n,n'>n}^N \varphi_{nn'} X_n X_{n'}},\label{eqMain}
\end{equation}
with $X_n$ a Pauli-$x$ operator acting on the $n$th qubit. 

Such gates, with arbitrary control of $\varphi_{nn'}$, together with local single-qubit rotations, form a universal gate set. Moreover, these gates have been shown to be advantageous in improving the performance and reducing circuit depth over a wide 
variety of quantum computation applications, to mention a few: Improved thresholds for quantum error correction, improved performance of optimization problems, reduced depth for the quantum Fourier transform, quantum volume circuits, Clifford circuits, $N$-qubit Toffoli gates, as well as for a wide range of random unitaries \cite{schwerdt2022comparing,grzesiak2020efficient,nemirovsky2025efficient,bravyi2022constant,liu2025performance,bassler2023synthesis,hoyer2003quantum,foxman2025randomu,galicia2020enhanced,nemirovsky2025phase,nemirovsky2025optimal}.

In recent years, several frameworks have been proposed for engineering entangling gates in large trapped-ion crystals, all making use of additional drive parameters, such as temporal or frequency-domain modulation of the gate's driving field's amplitude \cite{grzesiak2020efficient,duwe2022numerical} or phase \cite{kang2023designing,milne2020phase,lu2019global}, to address the quadratically growing constraints of achieving a target entangling map, $\varphi_{nn'}$, while minimizing drive power. One common approach is to focus on a small subset or a set of disconnected pairs of ions within a larger crystal, such that the number of quadratic constraints per ion reduces significantly and efficient solution methods become available \cite{choi2014optimal,blumel2021efficient,blumel2021power,leung2018robust,zhu2023pairwise,figgatt2019parallel}. Another line of work demonstrates the efficient construction of solutions for general entanglement maps, which are not necessarily power-optimal \cite{grzesiak2020efficient}. Finally, some studies employ custom or standard gradient-based minimization algorithms to obtain optimized solutions under the full set of constraints \cite{kosut2013robust,lu2019global,kang2021batch,gale2020optimized,duwe2022numerical}.

Our contribution here is twofold. First we introduce a toolbox, coined \textit{large-scale fast} (LSF), for designing multiqubit gates of the form of Eq.~\eqref{eqMain} for qubits based on ion crystals of up to 100s of qubits. Some of these tools can work in conjugation and enrich the previous methods above, while others are unique to LSF. Second, we make use of these techniques to investigate the physical properties of these gates, for gates of up to $N=100$ ions. Particularly, we show that while the multiqubit gates implement $\mathcal{O}\leri{N^2}$ entanglement operations, their minimal duration, $T_{\text{min}}$, below which general high-fidelity gate solutions cannot be found, scales linearly in $N$, constituting a scaling advantage. Furthermore, we derive and demonstrate an estimate of the power required to drive general gates in long crystals, and show it is directly deduced from the entanglement map, $\varphi$. Lastly, we introduce common error sources, e.g. motional modes frequency drifts, whose influence can be modeled within the physical framework used to derive LSF, and investigate how these errors scale the properties of the ion crystal and gate drive, providing design and calibration tolerances for large-scale trapped ions processors. 

The remainder of the paper is ordered as follows: in section~\ref{sec: formulation} we introduce the physical framework, formulate the gate-design problem and describe operational aspects of LSF; in section~\ref{sec: physical} we highlight the physical limits and scaling of laser power and $T_\text{min}$ with the number of ions and entanglement structure complexity; in section~\ref{sec: error} we perform an error analysis survey of common error mechanisms and describe their scaling with gate-time and number of ions; in section~\ref{sec: example} we demonstrate other physical properties LSF solution through a specific stabilizer-code gate example in a 49-ion chain. The next sections go into the details of the LSF method: in section~\ref{sec: derivation}, we describe the derivation of the LSF method; and in section~\ref{sec: benchmark} we describe the result of a benchmark test comparing the time-to-solution between our algorithm and competing common-use algorithms.

\section{Formulation of the gate design problem}\label{sec: formulation}

Trapped ion quantum computers use the normal-modes of motion of the ion crystal as a phonon bus, which mediates interactions between the ion qubits. This is performed by driving the motional sidebands of the ion crystal, which generates spin-dependent forces. A canonical example of this method is the M\o{}lmer-S\o{}rensen (MS) gate \cite{sorensen1999quantum,sorensen2000entanglement}. 
In recent years there have been many proposals and demonstrations which were focused on improving the utility and fidelity of MS gates. These methods are, at large, based on modulating the spin-dependent displacement forces by applying a non-static gate drive $\Omega_n(t)$ . While various signal representations may be used, here we analyze the problem in the spectral domain \cite{shapira2018robust,webb2018resilient,wang2022fast,blumel2023molmer}. That is, we use the Fourier representation over the frequency grid, that is symmetrically detuned around the qubit transition frequency, $\omega_0$ and spans the frequencies, $\omega_m=(m_0+m)\frac{\pi}{T_g}$, with $T_g$ the designated gate time, and $m$ and $m_0$ integers. The drive acting on the $n$th qubit is expressed as:
\begin{equation}
    \Omega_{n}(t)=\Omega\sin\leri{\omega_0 t}\left[\sum_{m}r_{nm}^{(c)} \cos(\omega_mt)+r_{nm}^{(s)} \sin(\omega_mt)\right],
\label{eqDriveExplicit}\end{equation}
with $r_{nm}^{\leri{c/s}}$ corresponding to the amplitude of tone $\omega_m$ driving ion $n$ at the cosine or sine quadratures and $\Omega$ a characteristic Rabi frequency. The spectral approach taken here is general in terms of the possible qubit drives, and offers conceptual advantages over alternative representations, as it is easier to manipulate analytically \cite{shapira2022robust} and provides physical intuition. Our approach is relevant to any form of qubit encoding; e.g. ground state, optical or metastable qubits \cite{allcock2021omg} as well as mixed-species crystals, qubit drive; e.g. Raman, optical or laser-free, and architecture; e.g. global beam \cite{wright2019benchmarking,manovitz2022trapped} or individually controlled ions.

Applying such drive results in an interaction Hamiltonian of the form  ($\hbar=1$),
\begin{equation}
\begin{split}
    H_\text{I}=&\Omega \sum_{n=1}^N X_n\sum_{j=1}^N\eta_j O_j^{\leri{n}}\leri{a_j^\dagger e^{i\nu_j t}+\text{H.c}}\\&\times \sum_{m=1}^M\left[r_{nm}^{(c)} \cos(\omega_mt)+r_{nm}^{(s)} \sin(\omega_mt)\right]
\end{split}
    \label{eqHI},
\end{equation}
with $a_j$ a bosonic annihilation operator of the $j$th phonon normal-mode of motion of the ion-crystal, at frequency $\nu_j$. The ion-phonon coupling is $\eta_j O_j^{\leri{n}}$, with $\eta_j$ the Lamb-Dicke parameter of the $j$th mode of motion and $O$ an orthonormal participation matrix such that $O_j^{\leri{n}}$ is the participation of the $n$th ion in the $j$th mode of motion.

Various assumptions and approximations are used in order to derive the interaction in Eq.~\eqref{eqHI} (given in full detail in Appendix~\ref{sec: app model}), most importantly the Lamb-Dicke approximation, i.e. assuming that $\eta^2_j(2n+1)\ll1$, and a rotating wave approximation used to omit unwanted coupling to the qubit carrier transitions, i.e. $\Omega_n\ll\nu$. Some of these approximations can be made more accurate by appropriately constraining the qubit modulations, $\Omega_n(t)$ \cite{shapira2018robust,shapira2020theory,webb2018resilient} (see ppendix~\ref{sec: app linear}).

The evolution due to $H_\text{I}$ is analytically solvable, yielding  a combination of qubit-dependent displacements of the phonon mode and qubit-qubit entanglement,
\begin{equation}
    U_\text{I}\leri{t}=\prod_{j=1}^N D_j\leri{\sum_{n=1}^N X_n\alpha_j^{\leri{n}}\leri{t}} e^{i\sum\varphi_{nn^\prime} (t)X_n X_{n^\prime}},\label{eqUI}
\end{equation}
with $D_j$ a displacement operator of the $j$th phonon mode. The displacements, $\alpha_j^{\leri{n}}$, and entanglement phases, $\varphi$, are directly computed from the system parameters $\eta_j$, $O_j^{\leri{n}}$, $\nu_j$, and the qubits drives, $\Omega_n$ (see Appendix~\ref{sec: app model}). 

The evolution in Eq.~\eqref{eqUI} gives rise to the gate design problem, namely obtaining the set of qubit drives, $\left\{\Omega_n\right\}_{n=1}^N$ such that at the gate time, $t=T_g$, all the displacements vanish, $\alpha_j^{\leri{n}}=0$ and such that $\varphi_{nn^\prime} (T_g)$ take predetermined values, $\varphi_{nn'}^\text{(ideal)}$, corresponding to a desired entangling gate. Gate design solutions which make use of a minimal driving power are typically preferable: they are better aligned with the approximations used to derive $H_\text{I}$; some sources of error scale with the drive power; and physical realizations are often limited by the drive power. 

Thus, the gate design problem becomes a norm-minimization problem, under quadratic and linear constraints:
\begin{equation}\label{eq 6: problem in r}
\hat{r} = \arg\min_{r} \|r\| \quad \text{subject to}\quad  \left\{\begin{array}{ll}\alpha_j^{(n)}(T_g) = 0&\forall j,n\\ \varphi(T_g) = \varphi_{nn'}^\text{(ideal)}&\forall n,n^\prime\end{array}\right.,
\end{equation}
with $j,n,n^\prime=1,...,N$. Crucially, the $\alpha_j^{\leri{n}}$s scale linearly in the $\Omega_n$s. Additional constraints, typically also linear in the $\Omega_n$s may be imposed in order to improve the accuracy of $H_\text{I}$ and to improve the gate's resilience to various sources of error and noise, e.g. calibration errors of the $\nu_j$s (see Appendix~\ref{sec: app linear}).  

The entanglement phases, $\varphi_{nn'}$ scale quadratically in the $\Omega_n$s, giving rise to NP-hard constraints \cite{Garey1979Computers,Blume1989Theory}. These constraints can be arranged into a linear constraint matrix $L$ with kernel $K$ and a set of real-valued $M\times M$ symmetric quadratic constraint matrices $A_{nn'}$. The form of the $A_{nn^\prime}$s arises naturally in the derivation of $U_\text{I}$ in Eq.~\eqref{eqUI} (see Appendix~\ref{sec: app model}).

A simpler representation of Eq.~\eqref{eq 6: problem in r} follows from the restriction of the drive into the space spanned by $K$,  by applying the transformation  $A_{nn^\prime}\to\mathcal {A}_{nn^\prime}=K^TA_{nn^\prime}K $. The optimization problem can be redefined as:
\begin{equation}
\hat{\boldsymbol{\mathcal{R}}} = \arg\min_{\boldsymbol{\mathcal{R}}} \|\boldsymbol{\mathcal{R}}\| \quad \text{subject to} \quad \boldsymbol{\mathcal{R}}^T\mathcal{A}_{s}\boldsymbol{\mathcal{R}}=\varphi_{s}^\text{(ideal)}\label{eqOptR}
\end{equation}
With $s=1,...,\mathcal{N}_c$ an ordering of the double index $nn^\prime$, $\mathcal{N}_c= \binom{N}{2}$ is the number of two-qubit pairings and $\boldsymbol{\mathcal{R}}^T=\begin{pmatrix}\boldsymbol{R}_{1}^{T} &\boldsymbol{R}_{2}^{T} & \cdots & \boldsymbol{R}_{N}^{T}\end{pmatrix}$ defined such that $\left(\boldsymbol{R}_{n}\right)_l$, is the amplitude associated with the $l$th vector in $K$ driving the $n$th ion. $\mathcal{A}_{s}$ are the set of $N\tilde{M}\times N\tilde{M}$ symmetric sparse matrices, written such that $\boldsymbol{\mathcal{R}}^{T}\mathcal{A}_{s}\boldsymbol{\mathcal{R}}=\boldsymbol{R}_n^{T}\mathcal{A}_{nn'}\boldsymbol{R}_{n'}$ and $\tilde{M}=\dim{K}$. 

The norm in Eq.~\eqref{eqOptR} used to evaluate $\left|\boldsymbol{\mathcal{R}}\right|$ is intentionally left undefined, as different physical realizations correspond to different choices of norm. In what follows bellow, we make use of the standard Euclidean norm. 

By using the `$\boldsymbol{\mathcal{R}}$' notation the gate design problem assumes the form of that described in Ref. \cite{shapira2020theory}, for homogeneously driven ions. Other realization of drive configuration, such as beams that illuminate sections of the ion-chain also conform to the form of Eq.~\eqref{eqOptR} \cite{solomons2025programmablequantumcomputingtrappedions}.

In the adiabatic limit, i.e. for a gate time $T_\text{g}$ much larger then the inverse motional modes density, the $A_{nn^\prime}$s become approximately diagonal and satisfying the quadratic constraints is trivial (see Appendix~\ref{sec: app adiabatic}). However due to the crowding of the mode frequencies in large ion crystals this results in impractically slow gates. Indeed as $2\pi\Delta\bar{\nu}\cdot  T_\text{g}\rightarrow1$, with $\Delta\bar{\nu}$ the average spacing of motional mode frequencies, the tones of the driving field strongly interact with many modes and the coupling matrices are in general dense, making the optimization problem non-trivial.

Before continuing to an analysis of multiqubit gates, we briefly give an account of operational aspects of LSF (followed by a detailed account in sections~\ref{sec: derivation} and~\ref{sec: benchmark}). LSF operates by transforming a special seed solution of the gate-design problem, coined a \textit{zero phase solution} (ZPS), into an approximate solution of the optimization problem in Eq.~\eqref{eqOptR}, with a polynomial-time process that has two main parts - a step that converts the ZPS to a solution satisfying the quadratic constraint, and then a set of norm-reduction steps. This process can be performed with several seed ZPSs. Thus, the `hardness' of the problem is transformed to the task of generating a finite set of ZPSs, which might be inefficient, after which generating a solution or each ideal target, $\varphi_s^\text{(ideal)}$, is done efficiently.

\section{Scalability in large ion crystals} \label{sec: physical}
We use LSF in order to investigate the performance and scaling laws of multiqubit entanglement operations in large ion crystals. To this end we consider many trapped ions systems which vary in number of ions, gate duration, entanglement target, and drive spectra, among other examples. \\
First, we consider the total Rabi frequency required to drive an entanglement gate, which we quantify by $\left|\boldsymbol{r}\right|\equiv\sqrt{\sum_{n,m}\leri{\boldsymbol{r}_n}_{m}^{2}}$. Since the gate design stems from an NP-hard problem, one would expect that predicting the required total Rabi frequency, before solving the optimization problem, to be challenging. Nevertheless such a prediction is useful as it can inform circuit construction and as an a-priori stopping criteria for the norm-reducing process. 

As a benchmark we recall the expression for the Rabi frequency which is required to drive a single mode MS gate, $\Omega_\text{MS}=\frac{\sqrt{\left|\varphi_\text{MS}\right|}}{\sqrt{2\pi}\eta T}$,  where $\varphi_\text{MS}$ is the entanglement phase, and $\eta\propto\nu^{-1/2}$ is the Lamb-Dicke parameter corresponding to the mode of motion at frequency $\nu$. This expression is valid in the adiabatic limit, i.e. for $T_\text{g}\gg\nu^{-1},\Delta\nu^{-1}$. We then generalize $\Omega_\text{MS}$  to a system in which the target $\varphi_{nn'}$ is native, i.e. it can be implemented with a global driving field. This is achieved in the case where that the normal-modes of motion of the ion crystal are the eigenvectors of the matrix $\varphi_{nn'}$ \cite{shapira2020theory}, and each mode accumulates a phase that is the corresponding eigenvalue. We incorporate this change by replacing $\varphi_\text{MS}\mapsto\sum_{j=1}^N\left|\varphi_j\right|$ , where $\left\{\varphi_j\right\}$ are the eigenvalues of $\varphi_{nn'}$. This sum is known as the \textit{nuclear norm} of the matrix $\varphi_{nn'}$. Furthermore we replace $\eta$ with the average $\eta$ over all modes of motion. When we go beyond native targets, we replace all the elements of $\varphi_{nn'}$ with their absolute value. Lastly, for independently driven ions we expect the power to scale linearly with $N$. Thus, we conjecture an estimate, $\Omega_\text{nuc}$,
\begin{equation}
    \Omega_\text{nuc}=k\frac{\sqrt{N \|\varphi\|_\text{nuc}}}{\sqrt{2\pi}\langle\eta\rangle T_\text{g}},\label{eqNuclearNorm}
\end{equation}
with $\|\cdot\|_\text{nuc}$ the nuclear norm of a matrix with its elements taken in absolute value, and $k$ a constant which depends on the choice of implemented linear constraints (e.g. additional linear constraints ensuring gate robustness). 

\begin{figure}
\centering{}\includegraphics[width=1\columnwidth]{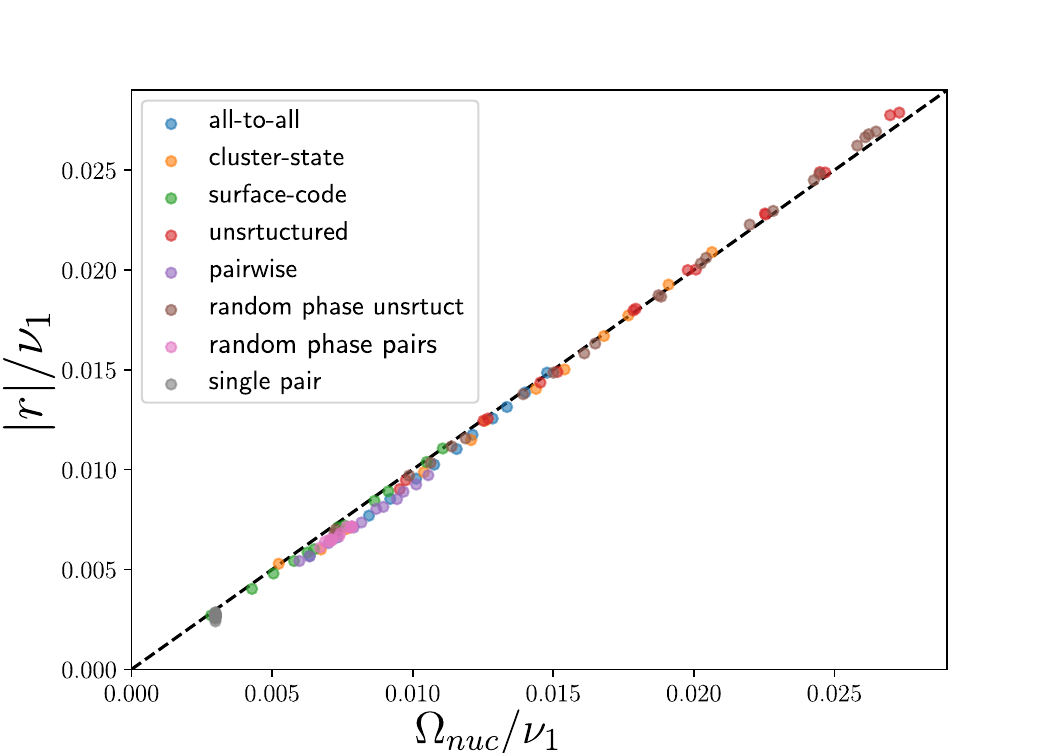} \caption{Correlation between the solution drive amplitude and its estimator $\Omega_{\text{nuc}}$ for 150 different phase maps on a single $N = 50$ ion crystal. Color indicates the category of the multiqubit coupling maps. These include \textit{structured maps} (e.g., all-to-all), where $N$ or a subset of ions follow a defined entangling pattern, and \textit{unstructured maps}, where phases are randomly assigned to ion pairs. Phases are either restricted to $\{0, \pi/4\}$ or drawn from the continuous range $[-\pi/4, \pi/4]$ for \textit{random-phase} maps. The dashed line indicates perfect correlation, $|r| = \Omega_{\text{nuc}}$.}
\label{figNuclear}
\end{figure}
To test the predictive power of $\Omega_\text{nuc}$, we generated a large set of entanglement maps for an equidistant crystal of 50 trapped $^{40}\text{Ca}^+$ ions, with an ion spacing of $d=5\mu\text{m}$ and a fixed gate time of $T_g = 780\mu\text{s}$, that drive the crystals transversal (radial) modes of motion. These maps are grouped into several categories - structured maps include configurations where the entangling pattern follows a defined geometry or purpose, such as \textit{all-to-all}, \textit{cluster-state}, \textit{surface-code}, \textit{pairwise}, and \textit{single pair} couplings. In these maps, the entangling phase between specific pairs is set to either $0$ or $\pi/4$, according to the group pattern. Moreover, we draw subsets of size $N'<50$ and repeat the gate with the pattern applied only to the ions in the subset. In contrast, \textit{unstructured} maps are generated by assigning a non-trivial entangling phases to randomly selected pairs. For some groups we drop the $0$ or maximal-entangling phase assertion and assign different pairs a randomly drawn phase within the range $[-\pi/4, \pi/4]$. For each entanglement target we generated solutions from five different ZPS's and averaged the resulting $\left|\boldsymbol{r}\right|$ values. Here and in the results below, solutions to the optimization problem are obtained such that the total deviation from the quadratic constraints, $\sum_s\left(\varphi_s-\varphi_s^{(\text{ideal})}\right)^2$, is at most $10^{-4}$, which loosely captures the gate's infidelity. 

The results are summarized in Figure~\ref{figNuclear}, where each point corresponds to a specific $\varphi_{nn'}$ example, and color is indicative of the entangling map category. For each such point the resulting drive amplitude, $\left|\boldsymbol{r}\right|$, is compared with the prediction $\Omega_\text{nuc}$. Remarkably, the data collapse shown in Fig.~\ref{figNuclear} implies that the details of the mode structure and frequency are largely irrelevant to the required Rabi frequency. The analogy to a globally driven system provides an intuitive explanation: Driving the ions independently is equivalent to an effective `reshaping' of the participation of each ion in each of the normal-modes, such that the resulting reshaped mode participation map fits the required spin coupling map. While this picture is intuitive, it could not have been verified without the ability to compute optimal, multi-mode, entanglement gates on large ion-crystals, afforded by LSF.

The factor $k$, empirically set to be $0.78$ from a fit to the data, is applicable when only the bare gate constraints are applied. With the inclusion of additional constraints, that aim at robustifying the gate against noise (e.g. reduce gate sensitivity to changes in mode frequencies), this factor assumes larger values, reflecting the additional cost associated with the inclusion of more conditions , see Appendix~\ref{sec: app nuc linear}. 

It should be noted that our estimation is heuristic, specifically it assumes that the modes of motion of the ion crystal are global, i.e. that in general all ions are coupled to each other. Indeed, it is easy to come up with cases in which ions vibrate independently from each other and ion-ion entanglement is impossible, e.g. Refs. \cite{schwerdt2024scalable,moses2023race}, yet the nuclear norm estimate will not diverge. 

We remark that while the eigenvalues of $\varphi$ play a crucial role in determining the gate drive power, they trivially also carry information about the amount of entanglement generated by the gate. Indeed, for small values of $\varphi$ the average single-qubit Von-Neumann entropy scales as the Frobenius norm squared of $\varphi$, i.e. the sum over its eigenvalues squared (see Appendix~\ref{sec: app von}), $\overline{S}_\text{VN}\leri{\varphi}\approx\sum_{n}\varphi_{n}^2= \|\varphi\|_\text{F}^2$. For these norms we have the bound, $\|\varphi\|_\text{F}\leq\|\varphi\|_\text{nuc}$, showing that the drive power puts a bounds on the average entanglement entropy generated by the gate.

\begin{figure}[!t]
\centering
\includegraphics[width=0.9\columnwidth]{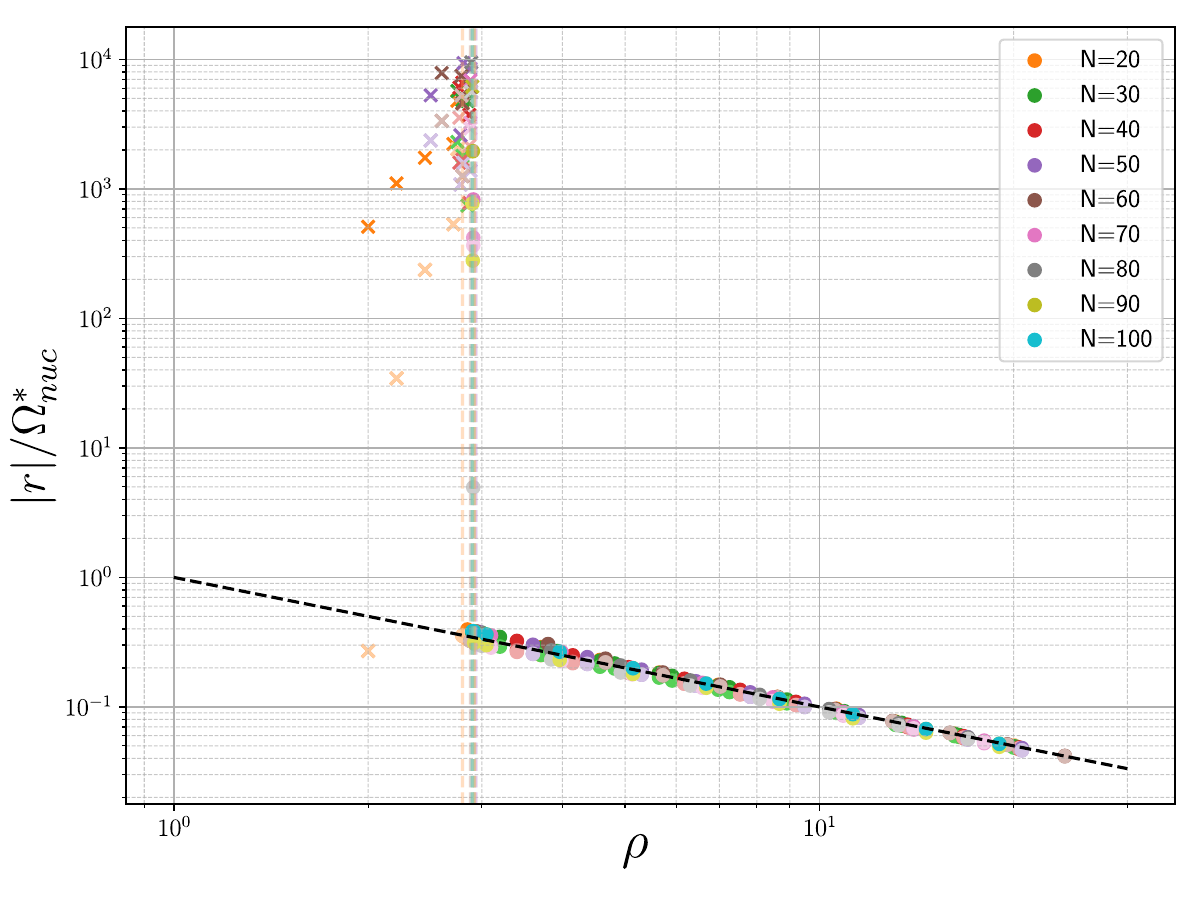} 
\caption{Summary of results obtained by our method, shown as total required Rabi frequency vs. gate time for various lengths of ion crystals (color). Each point corresponds to a different entanglement operation. The gate time (horizontal) is normalized by $T_\nu$, and the total Rabi frequency, $\left|r\right|$, (vertical, log scale), normalized according to the nuclear-norm based estimation, $\Omega_\text{nuc}$ and $T_g/T_\nu$. As seen with this scaling, all our solutions collapse on a single line (black dashed), signifying an inverse proportion between required power and gate time.}
\label{figFast}
\end{figure}

Following the validation of Eq.~\eqref{eqNuclearNorm} for a single ion-crystal and specific gate time, we proceeded to a comparison between systems of differing ion-count and gate-times. With our choice of harmonic-basis to the gate drive, faster gates contain less frequency components in the vicinity of mode frequencies, that resolve the crystal motional modes. This naturally leads to a controllability limit over the individual phases acquired by different modes, and consequently limits the accessible entangling map space.  Indeed, in the single-tone, far-detuned regime (\( \delta \gg \Delta \nu_j \)), a nearly uniform detuning from all motional modes gives rise to localized phonon wave packets and correspondingly local entangling interactions between ions~\cite{kyprianidis2024interaction,britton2012engineered,porras2006quantum}. A natural timescale to separate the different regimes is the average mode density,
\begin{equation}
    T_{\nu}= \frac{N\pi}{ \sum_j \Delta\nu_j} \equiv \pi(\Delta\bar{\nu})^{-1},
\end{equation}
proportional to the average reciprocal group velocity $v_g^{-1}(\omega)$~\cite{savill2025high} of phonons in the ion-crystal. As the mode density grows linearly, on average, with the crystal size, so does $T_\nu$. We define the dimensionless quantity  $\rho=T_g/T_\nu$ as the ratio between the gate time and the average mode-density time-scale. This dimensionless parameter provides a consistent framework for comparing systems of different sizes. The drive amplitude is rescaled by $\Omega_{\text{nuc}}^*=\Omega_{\text{nuc}}\cdot\rho$, which replaces the explicit gate-time dependence in Eq.~\eqref{eqNuclearNorm}, with the characteristic time $T_\nu$. $\Omega_{\text{nuc}}^*$ is a fixed amplitude estimator per ion-crystal.\\
This universal scaling is demonstrated in Figure~\ref{figFast}, where results from ion crystals with \( N = 20 \) to \( N = 60 \) collapse onto a single curve when plotted as a function of $\rho$. Within each color group (indicating system size), each point represents a specific entangling operation and gate duration, averaged over five optimization runs. From Eq.~\eqref{eqNuclearNorm} we expect the normalized gate amplitude to scale as $\rho^{-1}$, shown by the unity slope dashed black line. The data collapse provides compelling evidence for a non-trivial property: given a physical description of the ion crystal, a target gate time, and a desired entangling operation, one can estimate the required drive amplitude without solving the gate-design problem in full.

Another emergent universal feature is the threshold below which no feasible solution can be found. These breakdown points, marked by vertical dashed lines, cluster around \( \rho = 2.9 \). Attempts to design gates of shorter durations consistently fail: these solutions require significantly higher drive power then anticipated from $\Omega_{\text{nuc}}$, and many exceed the accepted gate error threshold (marked by ×). This signifies a minimal gate time  $T_{\text{min}}\simeq2.9\cdot T_\nu$ which scales linearly with the crystal size. It is worth noting that $T_{\text{min}}$ is exemplified here only using dense entangling maps involving all ions. Gates faster then $T_{\text{min}}$ do exist for entangling maps that couple only a part of the crystal \cite{wang2022fast,savill2025high}.
\section{Error scaling}\label{sec: error}
Beyond the scaling of resources with crystal size, an equally important consideration is how tolerable noise levels scale for gates involving many ions. To address this, we apply our formalism to analyze the impact of three dominant error mechanisms: drifts in the motional-mode frequencies, drive amplitude fluctuations, and motional heating, and quantify their respective contributions to gate infidelity. These processes are well known as dominant contributions in the noise budget of trapped-ions entanglement gate demonstrations \cite{ballance2016high,lotshaw2023modeling}. Here, we quantify their impact in crystals of tens of ions. These sources of error remain within the analytic framework of LSF, and can be modeled as modifications of the quadratic and linear matrices that determine mode displacements and entangling phases at gate time, or of the gate drive itself. Errors that are beyond the analytic model, e.g. higher-order Lamb-Dicke terms, are a subject of future research.

For small error terms and modes temperature $\bar{n}_j\ll1$, the total gate infidelity, $\mathcal{E}=1-F$, with $F$ the gate fidelity, can be expressed as a sum of displacement and phase errors \cite{kang2023designing,bentley2020numeric}:
\begin{equation}	
\mathcal{E}\approx \mathcal{E}_\alpha + \mathcal{E}_\varphi =\sum_{n,j}\frac{1}{4}\|\alpha_j^{(n)}\leri{T_\text{g}}\|^2 +\sum_{n,n'}\Delta\varphi_{n,n'}^2, \label{eqGateError}
\end{equation}
with $\Delta\varphi_{n,n'}=\varphi_{n,n'}\leri{T_\text{g}}-\varphi_{n,n'}^{\leri{\text{ideal}}}\leri{T_\text{g}}$.

\textbf{Motional-mode frequency drifts}: In the case of radial modes, such drifts can arise from shifts in the resonance frequency of the Paul trap RF resonator. The single-ion radial secular frequency scales as
$\nu_x \sim V_{\text{RF}}$, with $V_{\text{RF}}$ the voltage of the oscillating field that generates the trapping pseudo-potential. A drift in this voltage, $\Delta V$, shifts all radial mode frequencies according to $\nu_x^j\to\nu_x^j+\Delta V\cdot\delta_x^j$ with $\delta_x^j$ susceptibilities with a weak dependence on $j$. The resulting phase and displacement errors can be estimated by recalculating the integrals in Eqs. \eqref{eqUnitaryDisplacement}  and \eqref{eqUnitaryEntanglement} with the shifted frequencies.
\begin{figure}
    \centering
    \includegraphics[width=1.0\columnwidth]{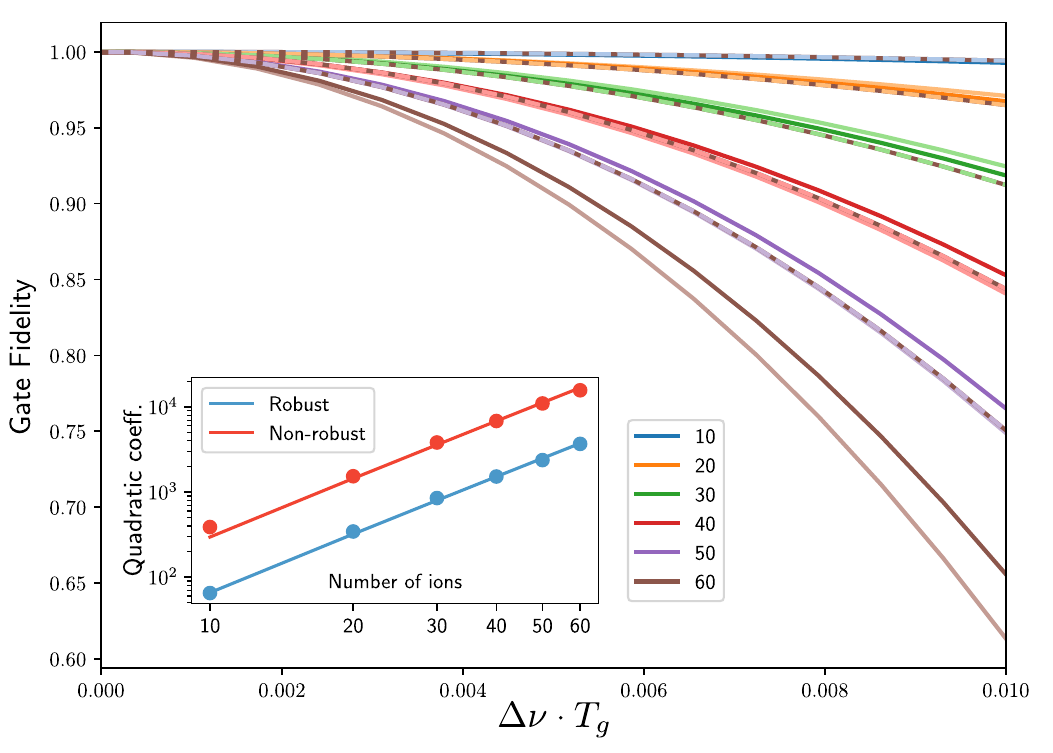} 
    \caption{Gate fidelity, $F$, due to a drift in the trap RF voltage, estimated for gates generating GHZ states for ion-crystals of 10 to 60 ions. Gate times of $T_g=2\cdot T_{\text{min}}$ (light) and $3\cdot T_{\text{min}}$ (dark) collapse to similar trends when rescaling the frequency error by $T_g$ (horizontal). Dashed colored-brown lines show gates on 60 ions where only a subset (of size coded by the color) of the ions are entangled, forming a sub- crystal size GHZ state. Inset: A quadratic approximation of the infidelity, with  Filled circles showing the quadratic coefficient for varying crystal size, with (blue) and without (red) additional linear constraints aiming at reducing the gate sensitivity to such drifts. Both coefficients scale almost linearly with the number of bi-partite couplings in the gate, $N_\text{c}\sim N^2$, yielding a constant slope of approximately 2.}
    \label{figErrorRF}
\end{figure}
Figure~\ref{figErrorRF} illustrates these errors for multiple gate solutions, each designed to generate a GHZ state. The solutions are grouped by ion number, with colors ranging from blue (10 ions) to brown (60 ions). The error parameter $\Delta\nu$ is normalized by the gate time $T_g$, which facilitates the collapse of error curves with different gate times onto similar trends for a given crystal size. This is demonstrated by two solutions per crystal size: bright-colored curves correspond to faster gates with $T_g = 2T_{\text{min}}$, and dark-colored curves to slower gates with $T_g = 3T_{\text{min}}$. Brown dashed curves correspond to gates in the 60-ion chain where only a subset of ions participate (according to color). These error curves closely match those of fully-connected gates on smaller crystals of the same number of participating ions.

This data shows that the gate errors under this noise model scale quadratically in the frequency error,
\begin{equation}
\mathcal{E}_{\Delta\nu}\approx k N_\text{c}\left(\Delta \nu T_g  \right)^2.
\end{equation}
Here, $N_\text{c}=\tfrac{N(N-1)}{2}$ is the number of ion–ion couplings and $k$ is an additional ($N$- and $\Delta\nu$-independent) constant. This constant is fitted from the data to $k
\approx4.5$ (inset) showing the scaling of the quadratic coefficient with $N_\text{c}$ (red). The gate can be made robust to this type of error by including appropriate linear constraints, shown in Ref. \cite{shapira2018robust,webb2018resilient,shapira2020theory} and Appendix~\ref{sec: app linear}, which are shown to reduce  $k$ to $
\approx1.0$ (blue).

\textbf{Drive amplitude fluctuations}: For laser driven ions, such fluctuations can arise, e.g. from instabilities in the laser intensity or beam pointing. To model this effect, we rescale the spectrum driving ion $n$ as $r_n \to r_n(1+\epsilon)$, where $\epsilon$ denotes the relative amplitude change. Since the displacement integrals are linear in $r$, and solutions are limited to $K = \ker(L)$, this rescaling of $r_n$ does not cause displacement errors. Amplitude fluctuations, however, do cause phase errors that can be written as,
\begin{equation}
\begin{split}
    \mathcal{E}_{\varphi} 
    &= \sum_s \left( \mathcal{R}(1+\epsilon)\mathcal{A}_s\mathcal{R}(1+\epsilon)-\mathcal{R}\mathcal{A}_s\mathcal{R}\right)^2 \\
    &= \left(4\epsilon^2+4\epsilon^3+\epsilon^4\right)\sum_s\varphi_s^2 .
\end{split}
\label{eqErrorLaserAmplitude}
\end{equation}
For homogeneous stochastic noise $\epsilon\sim P_\epsilon$, the powers of $\epsilon$ in Eq.~\eqref{eqErrorLaserAmplitude} are replaced by the corresponding distribution moments. For Gaussian noise, $E[\epsilon^{2n}] = \sigma^{2n}(2n-1)!!$, yielding the expected error,
\[
\mathcal{E}_{\varphi} = \left(4\sigma^2+3\sigma^4\right)\sum_s\varphi_s^2 .
\]
\begin{figure}
    \centering
    \includegraphics[width=1.0\linewidth]{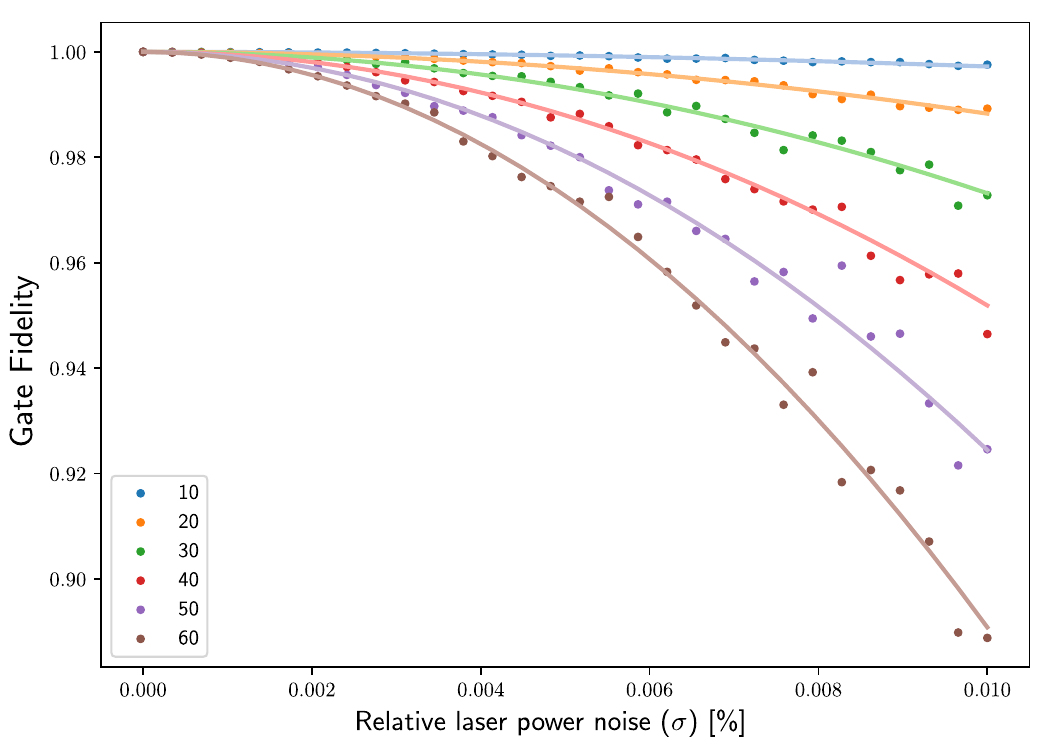}
    \caption{Gate infidelity due to fluctuating laser amplitude. For each crystal size, the drive amplitude was rescaled by $1+\epsilon$ with $\epsilon\sim\mathcal{N}(0,\sigma^2)$. Filled circles correspond to numerical averages (mean of 300 samples), while solid lines show the expectation value of Eq.~\eqref{eqErrorLaserAmplitude}.}
    \label{figLaserAmpError}
\end{figure}
Figure~\ref{figLaserAmpError} shows the expected error for crystals of size 10–60 ions. Filled circles represent averages over 300 Gaussian samples per variance $\sigma^2$, while solid curves follow the analytic prediction. For maximally entangling gates (e.g., Clifford multiqubit gates), and under the assumption of small noise variance, the error scales as
\begin{equation}
    \mathcal{E}_{\Delta r} \approx 2\pi\sigma^2 N_c ,
\end{equation}
with $N_c$ the number of nontrivial ion-ion couplings. Thus, the variance of the noise must decrease inversely with $N_c$, which grows, at most, quadratically with $N$.

\textbf{Motional heating:} Heating of secular ion motion is a common dephasing process that limits coherent control in ion-crystals. It typically increases with decreasing ion-electrode distance, and is more severe in room-temperature vs. cryogenic traps. Following the analysis of~\cite{he2023scaling}, the contribution of motional heating to gate error is bounded by
\begin{equation}
    \mathcal{E}_{\Gamma}\le \sum_{n,n',j}\Gamma_j\left\|\int_0^{T_g}\alpha_j^{*(n)}(t)\alpha_j^{(n')}(t)\,dt\right\|,\label{eqHeatingRateBound}
\end{equation}
where $\Gamma_j$ is the heating rate (in quanta per unit time) of mode $j$. The rate depends both on the electric noise power spectral density $S_E(\omega)$ and the spatial correlation of electric-field noise across the ions \cite{brownnutt2015ion},
\begin{equation}
    \Gamma_j=\frac{e^2}{4m\omega_j^2}\sum_{n,n'}O_j^{(n)}O_j^{(n')}S_{E}^{(n,n')}(\omega_j),\label{eqHeatingRateModel}
\end{equation}
with
\[
S_{E}^{(n,n')}(\omega)=\int d\tau\, e^{-i\omega\tau}\langle E^{(n)}(\tau)E^{(n')}(0)\rangle , 
\]
the cross-spectral noise density function. If the noise is perfectly correlated, $S_E^{(n,n')}(\omega)\equiv S_E(\omega)$, then only the center-of-mass (COM) mode contributes to the expression Eq.~\eqref{eqHeatingRateModel}. For a more realistic scenario, we consider a noise model with $S_E(\omega)\propto \omega^{-1}$ and exponentially decaying correlations, such that  $S_{E}^{(n,n')}(\omega)\propto \omega^{-1}e^{-|x_n-x_{n^\prime}|/\zeta}$, with $\zeta$ the correlation length of the ions at positions $x_n$.  
\begin{figure}
    \centering
    \includegraphics[width=1.0\linewidth]{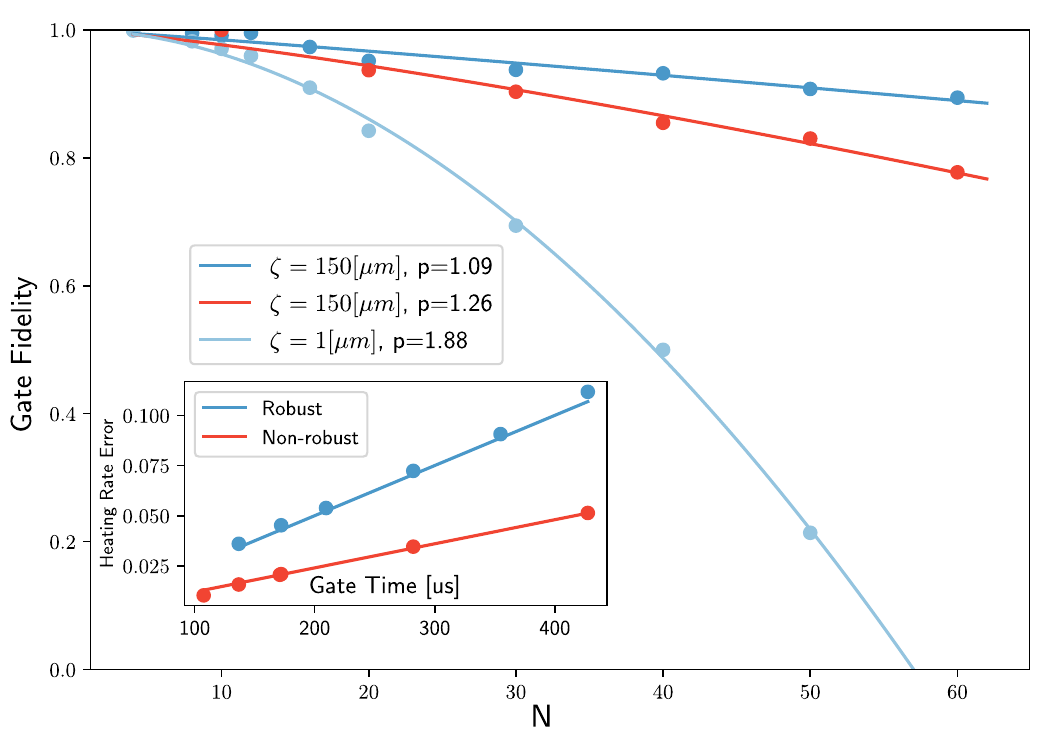}
    \caption{Infidelity due to motional heating. Main frame: Heating-induced errors for GHZ gates in crystals of 10–60 ions. To remove the trivial gate-time scaling, the COM-mode heating rate is normalized to $0.02$ quanta per gate time. Results are shown for correlated noise ($\zeta=150~\mu$m, spanning most of the crystal) and uncorrelated noise ($\zeta=1~\mu$m). Solutions including motional error minimization linear constraints (blue) are compared with unconstrained solutions (Red). Inset: Error growth with gate time in a 20-ion crystal. Errors increase linearly with $T_g$, as expected from Eq.~\eqref{eqHeatingRateBound}, with robust gates (blue) showing roughly half the error rate.}
    \label{figHeatError}
\end{figure}

Figure~\ref{figHeatError} shows the gate fidelity due to motional heating, for gates that generate GHZ states in ion-crystals of 10 to 60 ions. Since the minimal gate time increases with $N$, then in order to isolate the effect of noise correlations we rescale the COM mode heating rate to $0.02$ quanta per gate time. When noise is strongly correlated, $\zeta=150~\mu$m (blue, red), the fidelity decays roughly linearly with $N$. By contrast, for uncorrelated noise, $\zeta=1~\mu$m (bright blue), the fidelity decays quadratically in $N$. Although the sum in Eq.~\eqref{eqHeatingRateBound} runs over indices $n,n'=1,..,N$, not every ion couples equally to every mode in a given gate. As a result, the scaling is not purely quadratic (as would be expected if all modes contributed independently), but instead interpolates between linear and quadratic scaling depending on the correlation model. Similarly to motional-mode frequency drifts above, robust gate (blue) are compared to non-robust gates (red), and show a reduced sensitivity to heating-induced errors.

\section{Analysis of an example: Surface code stabilizer}\label{sec: example}
\begin{figure}
\centering{}\includegraphics[width=1\columnwidth]{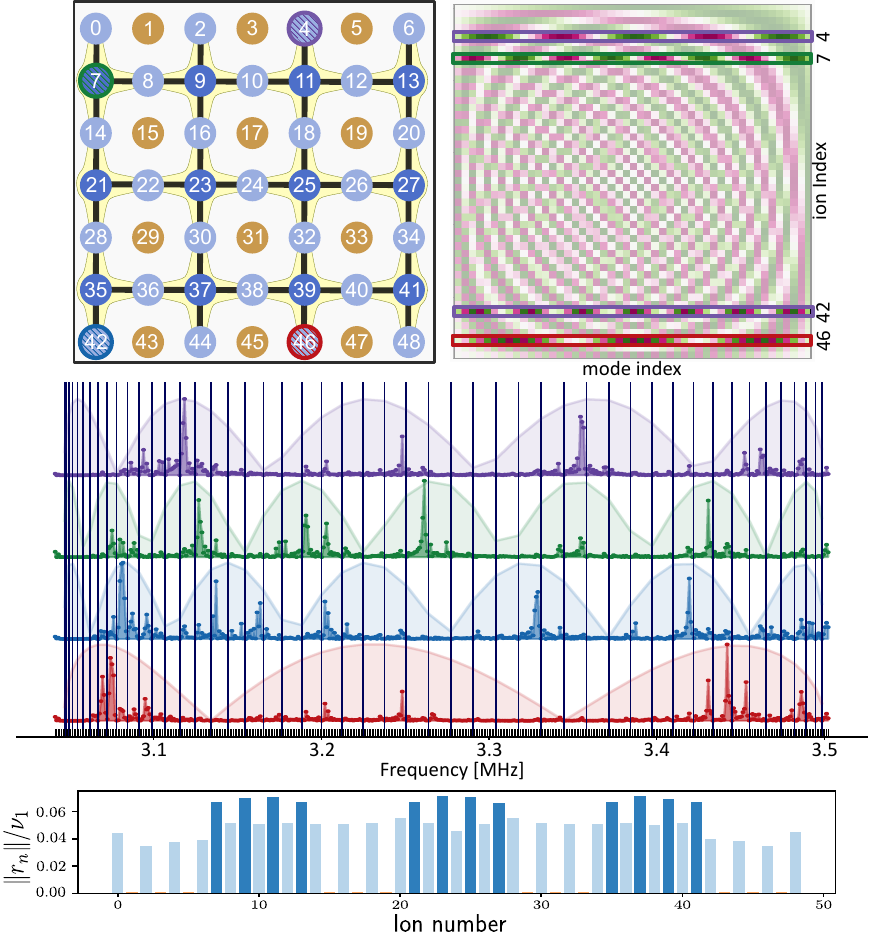} \caption{
Single-pulse surface-code stabilizer gate in a 49-ion crystal.
Top-left: Mapping of a linear 49-ion crystal onto a $7\times7$ square grid forming nine surface-code plaquettes. Dark-blue ions denote ancilla (node) qubits, light-blue ions form plaquette edges, and orange ions are uncoupled spectators. Upper-right: Identification of four representative participating ions used in the spectral analysis. Middle: Drive spectrum for these four ions during a single gate pulse. Vertical dark-blue lines indicate transverse motional-mode frequencies of the 49-ion crystal. Bottom: Corresponding ion-resolved drive amplitudes, showing higher power assigned to ancilla ions, reduced power on edge ions, and vanishing drive on non-participating ions
}
\label{figModel}
\end{figure}

We demonstrate solutions obtained by our method and highlight certain aspects of them via an example. Specifically, we outline the entanglement gate required for stabilizer measurements in surface codes. Here we consider a 49 ions crystal, entangling 33 ions using a single pulse. Figure~\ref{figModel} shows the formation of the stabilizer (top-left panel), with ions straightforwardly mapped into a $7\times7$ square grid, forming nine plaquettes that can be used to evaluate the $X$-parity of the plaquette vertices (a similar construction can be made for $Z$-parity stabilizers). In this mapping some ions remain uncoupled (orange) and can be used in subsequent operations; some form the edges of plaquettes (blue) and some are designated as ancilla qubits (dark blue). It was shown in Ref. \cite{singlegatestabilizer} that a surface code stabilizer measurement can be implemented with a single multi-qubit MS gate. Note, however, that our coupling map between ions involved in a stabilizer measurement is not all-to-all; rather it takes the form of a `cross'. This is in fact more efficient as it requires fewer non-zero entanglement phases. Indeed the nuclear norm of the cross coupling map is two times lower than that of the all-to-all coupling.  \par

Here we assume an equally spaced crystal of \({}^{40}\mathrm{Ca}^+\) ions with an inter-ion distance of \(5\,\mu\mathrm{m}\). Qubits are mapped onto the ground-state Zeeman \(5S_{1/2}\) manifold and are assumed to be driven with a \(400\,\mathrm{nm}\) laser field using a Raman transition, which couples the ions through transverse motional modes at frequencies between \(3\,\mathrm{MHz}\) and \(3.5\,\mathrm{MHz}\). We convert different ZPS's into solutions for this particular gate problem and compare their properties. The obtained solutions exhibit infidelity below \(1\times10^{-4}\), as defined in Eq.~\eqref{eqGateError}. The  middle panel of Fig.~\ref{figModel} provides an example of a drive spectrum obtained for four participating ions, labeled in the upper-right panel as 4 (purple), 7 (green), 42 (blue), and 46 (red). The dark-blue vertical lines correspond to motional-mode frequencies of the 49-ion crystal in one of the radial directions. The spectrum describes a pulse of duration $T\approx 640\,\mu\mathrm{s}$. Nonetheless, we obtain solutions for this problem setup for gate times down to \(T\approx 320\,\mu\mathrm{s}\). For comparison, to produce the same operation using sequential all-to-all two-qubit gates within $T\approx 640\,\mu\mathrm{s}$ requires that each two-qubit gate will have a duration of \(15\,\mu\mathrm{s}\). 

The obtained solution exhibits efficient use of the drive, manifested in two ways. First, the power assigned to each ion correlates with its participation in the gate. A representative drive amplitude solution is presented in the bottom panel of figure .~\ref{figModel}. Node ions (dark-blue) exhibit the highest drive amplitude, while edge ions (light-blue) are driven with lower amplitude, with a minor reduction for ions connected to a single qubit. The non-participating orange ions are trivially shut off.

A second property relates to how the total entangling phase is accumulated through all participating modes. Since the total entangling phase of an ion pair is weighted by the pair's participation in each mode (Eq.~\eqref{eqUnitaryEntanglement}), it is beneficial for the solver to entangle an ion pair through modes in which both ions participate substantially. Because there are \(\sim N^2/2\) pairwise constraints but only \(N\) modes, such efficient combinations are not guaranteed to exist. Nonetheless, the solver finds efficient patterns, as exemplified in the spectrum in Fig.~\ref{figModel}. The background color of the four spectra represents the participation of that ion in each mode, also summarized in the upper-right mode-participation matrix, where colored frames highlight the participation structure. In all cases, the solver assigns most power near modes where the ion participates substantially.
\begin{figure}
\centering{}\includegraphics[width=1\columnwidth]{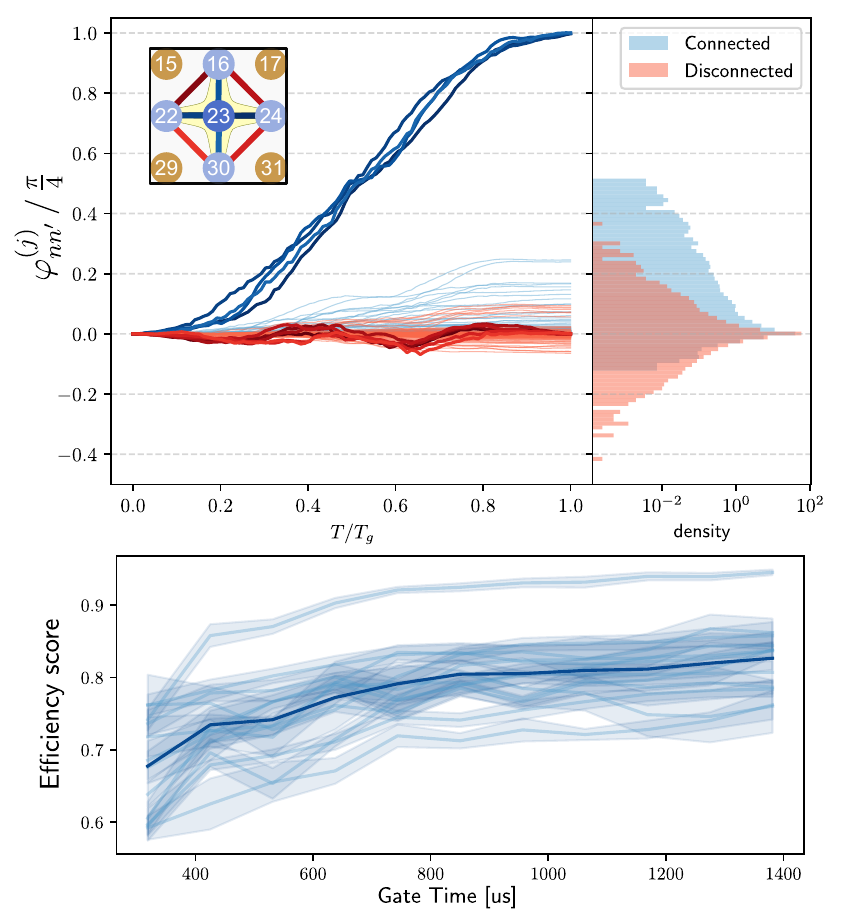} \caption{
Accumulation and suppression of entangling phases through shared motional modes.
Top-left: Time-resolved entangling phases for ion pairs along a selected node (highlighted in the inset). Blue traces show phases for connected ion pairs; red traces correspond to disconnected pairs. Darker curves indicate total entangling phase summed over all modes. Disconnected pairs acquire non-zero intermediate phases that cancel at the gate time.
Top-right: Histograms of pair-mode entangling phases collected from ten gate solutions, comparing connected (blue) and disconnected (red) ion pairs. Disconnected pairs exhibit a smaller spread, indicating efficient suppression of unwanted couplings despite shared mode participation. Bottom: Efficiency score, $S_{nn'}$, for connected ion pairs as a function of the total gate time. Thin traces correspond to individual connected links, with shaded regions indicating the spread across different optimized solutions. The bold curve shows the average link efficiency. }
\label{figEff}
\end{figure}

The factor-of-$N$ mismatch between the number of ion pairs and the number of motional modes implies that modes cannot be uniquely assigned to individual pairs. As a result, multiple ion pairs inevitably accumulate entangling phase through the same modes, generating unwanted cross-entanglement: a disconnected pair may acquire phase through a given mode even though it is not intended to be entangled. In typical solutions, this is mitigated by enforcing cancellation of the total entangling phase of disconnected pairs, rather than suppressing phase accumulation through each individual mode.

This mechanism is illustrated in the top-left panel of Fig.~\ref{figEff}, which shows the time-resolved entangling phases for ion pairs along a selected node (highlighted in the inset). Blue traces correspond to connected pairs and red traces to disconnected pairs; lighter curves show mode-resolved contributions, while darker curves represent the total phase summed over all modes. Although the total phase of disconnected pairs (dark red) is constrained to vanish at the gate time, it is composed of nonzero positive and negative contributions from different modes, resulting in oscillatory phase evolution during the gate.

From an efficiency standpoint, the solver suppresses these unwanted contributions by keeping the mode-resolved phases of disconnected pairs smaller than those of connected pairs. This is quantified by the histograms shown to the right of the panel, which collect pair-mode phase magnitudes from ten independent solutions. Disconnected pairs (red) exhibit a narrower distribution than connected pairs (blue), indicating that parasitic mode-mediated couplings are kept small. This reduces the fraction of the total drive power that must be allocated to cancel unwanted phases, improving overall gate efficiency.

The bottom panel of Fig.~\ref{figEff} shows how this efficiency degrades as the gate time is reduced. To quantify this effect, we define an efficiency score for each ion pair that measures how strongly its entangling phase is accumulated through modes in which both ions participate. The mode-participation weights, $w^{nn'}_j= \left|O_{j}^{(n)}O_{j}^{(n')} \right|$, are normalized to span $[0,1]$, 
\begin{equation}
\tilde{w}^{nn'}_j=\frac{w^{nn'}_j-\min\limits_j w^{nn'}_{j}}{\max\limits_j w^{nn'}_{j}-\min\limits_j w^{nn'}_{j}}
\end{equation}
and are used to define the efficiency,
\begin{equation}
\mathcal{S}_{nn'} = \frac{\sum_j \tilde{w}^{nn'}_j \, \lvert \varphi_{nn'}^{(j)} \rvert}{\|\varphi_{nn'}\|_1}.
\end{equation}
The efficiency scores of individual connected links are plotted as a function of gate time, with shaded regions indicating the spread across different optimized solutions, and the bold curve showing the average efficiency $\bar{\mathcal{S}}_{nn'}$. For long gate times, the efficiency remains approximately constant, indicating that the additional degrees of freedom available do not lead to more efficient mode usage. As the gate time is reduced, the efficiency decreases, reflecting suboptimal phase allocation due to reduced controllability. No solutions are found within the $10^{-4}$ tolerance for gate times below $T \approx 320\,\mu\mathrm{s}$, consistent with Fig.~\ref{figFast}.

\section{Derivation of the \textit{Large-scale fast} method} \label{sec: derivation}

In providing approximate solutions to Eq.~\eqref{eqOptR}, LSF takes two steps: First, solutions to the quadratic problem $\boldsymbol{\mathcal{R}}_\text{conv}$ are efficiently converted from a special seed vector $\boldsymbol{Z}$ called a ``zero phase solution''. $\boldsymbol{Z}$ is a non-trivial solution to a special case of the quadratic problem with $\varphi_{s}^{\leri{0}}=0$ for all $s=1,...,\mathcal{N}_\text{C}$. A set, $\{\boldsymbol{Z}\}$, is accumulated once and prior to the solution of any general $\varphi$. Second, the norm of $\boldsymbol{\mathcal{R}}_\text{conv}$ is minimized iteratively while adhering to the constraint $\varphi$ until an optimized solution $\boldsymbol{R}_\text{LSF}$ is reached. 
\subsection{Conversion of $Z$ into $R_\text{conv}$ }
A basic property of the zero-phase problem is that any solution $Z$ remains a valid solution when scaled by a real number $\lambda$. We exploit it to construct a `converted' solution, $\boldsymbol{R}_\text{conv} = \lambda\boldsymbol{Z}+\boldsymbol{D}$, with $\left|\boldsymbol{D}\right|$ assumed small, detailed below. For a large enough $\lambda$ small deviations from the ZPS generates arbitrarily large entanglement phases, that scale as $ \lambda \left| \boldsymbol{Z} \right|\left| \boldsymbol{D} \right|$. We linearize the quadratic constraints in the $\boldsymbol{D}$-vicinity of $\lambda\boldsymbol{Z}$ and obtain a set of $\mathcal{N}_\text{C}$ linear equations for $\boldsymbol{D}$,
\begin{equation}	
\varphi_{s}=2\lambda\boldsymbol{Z}^{T}\mathcal{A}_{s}\boldsymbol{D}+\mathcal{O}\leri{\epsilon},\label{eqNullToFull}
\end{equation}
with the assumption $\left|\boldsymbol{D}^{T}\mathcal{A}_{s}\boldsymbol{D}\right|\leq\epsilon$ for all $s=1,...,\mathcal{N}_\text{C}$ and $\epsilon$ is the acceptable gate error, which we typically specify to be $10^{-4}$.  The set of linear equations defined by Eq.~\eqref{eqNullToFull} yields the solution,
\begin{equation}	     
    \boldsymbol{R_\text{conv}}=\lambda\boldsymbol{Z}+\lambda^{-1}M_\text{pinv}^{-1}\boldsymbol{\varphi}\label{eqFQSol},
\end{equation}
with $M\in\mathbb{R}^{\mathcal{N}_\text{C}\times N\tilde{M}}$ a matrix whose elements are given by
 $M_{s,j}=2\left(\boldsymbol{Z}^{T}A_{s}\right)_{j}$ and $M_\text{pinv}^{-1}$ its pseudo-inverse. To ensure that the quadratic constraints are satisfied up to an accuracy $\epsilon$ we choose $\lambda$ that enforces the consistency condition,
\begin{equation}
\left|\left(\frac{1}{\lambda}M_{\text{pinv}}^{-1}\boldsymbol{\varphi}\right)^{T}\mathcal{A}_{s}\left(\frac{1}{\lambda}M_{\text{pinv}}^{-1}\boldsymbol{\varphi}\right)\right|\leq\epsilon,\label{eqConsistentLambda}
\end{equation}
for all $s=1,...,\mathcal{N}_\text{C}$.
\begin{figure}
\centering{}\includegraphics[width=1\columnwidth]{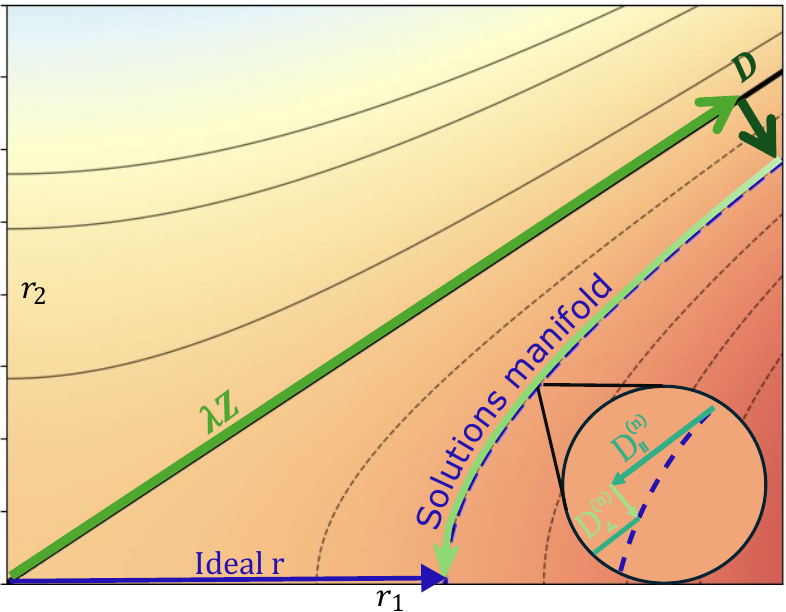} \caption{Simple example of the LSF method for a single quadratic constraint. We choose $\mathcal{N}_\text{C}=1$ and two tone amplitudes $r_1$ (horizontal axis) and $r_2$ (vertical axis), specifically $A_1=\text{diag}\left(1,-2\right)$ and $\varphi_1=2$. The constraint becomes a hyperbola (black dashed), and the colored contours show other possible values of $\varphi$. Our optimization rescales the zero-phase solutions, here by $\lambda=3$, and uses the high density of contours around $\lambda \boldsymbol{Z}$, (light green arrow), such that only a small excursion by $\boldsymbol{D}$ (bright green arrow) is required to convert $\boldsymbol{Z}$ to a solution satisfying the constraints, $\boldsymbol{R}_\text{conv}$. The locally ideal solution, $\boldsymbol{R}_\text{LSF}$ (blue arrow), is obtained by local minimization steps along the solution manifold, $\varphi_1=2$, composed of norm-reducing and error-reducing sub-steps (inset).}
\label{figNullToFull}
\end{figure}

Figure~\ref{figNullToFull} shows the conversion and optimization processes outlined above for the simple case of a single quadratic constraint, i.e. $\mathcal{N}_\text{C}=1$, and two amplitudes, $r_1$ and $r_2$. Specifically we use $\mathcal{A}_1=\text{diag}\left(1,-2\right)$ and $\varphi_1=2$, such that the constraints are satisfied along a hyperbola (dashed black) in terms of the two amplitudes, $r_1$ (horizontal axis) and $r_2$ (vertical axis). Zero-phase solutions extend from the origin to arbitrarily large amplitudes (dashed gray). We use a large zero-phase solution, $\lambda\boldsymbol{Z}$ (bright green arrow), such that a small deviation from it, $\boldsymbol{D}$ (dark green arrow), explores various values of $\varphi_1$. Indeed a converted solution, $\boldsymbol{\mathcal{R}}_\text{conv}$ (green arrow), generating the desired target is found close by. 
\subsection{Minimization of $\mathcal{R}_\text{conv}$ norm}
To reduce the solution norm, we define an iteration on $\boldsymbol{\mathcal{R}}$,
\begin{equation} 
\boldsymbol{\mathcal{R}}^{(n+1)}=\boldsymbol{\mathcal{R}}^{(n)}+\boldsymbol{{D}}^{(n)}.
\end{equation}
Here $\boldsymbol{\mathcal{R}}^{(0)}=\boldsymbol{\mathcal{R}}_{\text{conv}}$ is the converted ZPS. $\boldsymbol{D}^{(n)}$ is a correction term computed at each iteration and is built from two sub steps,
\begin{equation}
\boldsymbol{D}^{(n)}=\delta_{\parallel}\boldsymbol{D}_\parallel^{(n)}+\delta_{\perp}\boldsymbol{{D}}_\perp^{(n)},
\end{equation}
which are computed and applied in sequence. Particularly, $\boldsymbol{D}_\parallel^{(n)}$ is a \textit{norm reducing} step and $\boldsymbol{D}_\perp^{(n)}$ is an \textit{error reducing} step. The iteration terminates and $\boldsymbol{\mathcal{R}}^{(n)}=\boldsymbol{\mathcal{R}}^{\text{LSF}}$ if $\boldsymbol{D}^{(n)}$ is unable to significantly reduce the solution norm under the prescribed constraints.

To compute either correction $\boldsymbol{D}_{\text{x}}^{(n)}$ we linearize the constraints equations similarly to Eq.~\eqref{eqFQSol} to produce the following set of equations: 
\begin{equation}
\begin{split}\boldsymbol{\Theta}^{(n)}_\text{x}=
    \begin{pmatrix}\Delta\boldsymbol{\varphi}_s^{\left(n\right)}\\
\alpha_\text{x}
\end{pmatrix}=\begin{pmatrix}2{\boldsymbol{\mathcal{R}}^{(n)}}^\top\mathcal{A}_{s}\\
\boldsymbol{\mathcal{R}}^{\left(n\right)}
\end{pmatrix}\boldsymbol{D}_{\text{x}}\equiv\mathcal{M}^{(n)}\boldsymbol{D}_{\text{x}}\label{eqNormErrorStep}
\end{split}.
\end{equation}
with $\mathcal{M}^{(n)}\in\mathbb{R}^{\leri{\mathcal{N}_\text{C}+1}\times N\tilde{M}}$, and $\boldsymbol{\Theta}^{(n)}_\text{x}\in\mathbb{R}^{\leri{\mathcal{N}_\text{C}+1}}$, generalizations of $M$ and $\boldsymbol{\varphi}$ from Eq.~\eqref{eqFQSol} respectively. Furthermore, $\alpha_\text{x}$ controls the orientation $\boldsymbol{D}_\text{x}$. 

Performing the steps in sequence, first a \textit{norm reducing} step is derived by setting $\alpha_\parallel=-\left|\boldsymbol{\mathcal{R}}^{\left(n\right)}\right|$, enforcing a step in a direction minimizing the current solution norm. A solution to Eq.~\eqref{eqNormErrorStep} is found using the pseudo-inverse $\boldsymbol{D}_\parallel^{(n)}=\left[{\mathcal{M}_\text{pinv}^{(n)}}\right]^\top\boldsymbol{\Theta}^{(n)}_\parallel$, and the resulting constraint error is expressed as a polynomial in $\delta_\parallel$,
\begin{equation}
\begin{split}
    \varepsilon_{\varphi}^{(n,\parallel)}&=\sum_s \varepsilon_{\varphi,s}^{(n,\parallel)},\\
    \varepsilon_{\varphi,s}^{(n,\parallel)}&=\left(\varphi_s-\left(\boldsymbol{\mathcal{R}}^{(n)}+\delta_\parallel\boldsymbol{D}^{(n)}_{\parallel}\right)^\top
\mathcal{A}_{s}\left(\boldsymbol{\mathcal{R}}^{(n)}+\delta_\parallel\boldsymbol{D}^{(n)}_{\parallel}\right)\right)^2.
\end{split}\label{eqEpsilonPar}
\end{equation}
$\delta^{(n)}_\parallel$ is found by solving $\varepsilon_{\varphi}^{(n,\parallel)}(\delta_\parallel)=\epsilon_\text{n.r}$, providing the largest step size which yields an error equal or below $\epsilon_\text{n.r}$, an excess `norm-reducing` error term which is allowed to occur after the norm step. 
For the \textit{error-reducing} step, Eq.~\eqref{eqNormErrorStep} is re-applied, with $\boldsymbol{\mathcal{R}}^{(n)}$ replaced by $\boldsymbol{\mathcal{R}}^{(n)}+\delta_{\parallel}^{(n)}\boldsymbol{D}_\parallel^{(n)}$ , and $\alpha_\perp=0$, which guaranties the resulting norm of $\boldsymbol{\mathcal{R}}^{(n+1)}$ is only second-order sensitive to $\left|\delta_\perp \boldsymbol{D}_\perp^{(n)}\right|$.  This time, the resulting constraint error $\varepsilon_{\varphi}^{(n,\perp)}$ is minimized by solving $\partial_{\delta_\perp}\varepsilon_{\varphi}^{(n,\perp)}=0$, with $\varepsilon_{\varphi}^{(n,\perp)}$ defined similarly to Eq.~\eqref{eqEpsilonPar}. As both $\varepsilon_{\varphi}^{(n,\parallel)}$ and $\varepsilon_{\varphi}^{(n,\perp)}$ are fourth and third polynomials in $\delta_{\parallel/\perp}$, straightforward algebric methods, particularly Ferrari's and Cardano's methods \cite{ireland1990classical}, are applied to solve for $\delta^{(n)}_\parallel$ and $\delta^{(n)}_\perp$, respectively.

To accept the step we require the step error to be below the gate acceptable error rate, that is   $\varepsilon_{\varphi}^{(n,\perp)}<\epsilon$. If the error does not meet the gate error threshold, the step is discarded and a smaller norm-reducing step is calculated by decreasing $\epsilon_\text{n.r}$. Execution is halted  when either $\epsilon_\text{n.r}$ or the relative norm change $\left\|1-\frac{\boldsymbol{\mathcal{R}}^{(n+1)}}{\boldsymbol{\mathcal{R}}^{(n)}}\right\|$ are reduced below a predefined tolerance value.

\subsection{Aggregation of Zero Phase Solutions}
As our method relies on the existence of ZPS's, these should be obtained at the outset. Therefore, before solving any specific gate (given by  $\varphi^{(\text{ideal})}$), we first search for nontrivial solutions to the homogeneous problem $\boldsymbol{Z}\mathcal A_s\boldsymbol{Z}=0,\,\forall s$. \\
Zero phase solutions are produced in a similar iterative procedure. Beginning from a (pseudo-) random vector $\boldsymbol{z}_0$ of unit-length, a linear correction $\boldsymbol{\zeta}^{(n+1)}$ is obtained for $\boldsymbol{z}^{(n)}$ through the set of equations:
\begin{equation}
\begin{pmatrix}\boldsymbol{z}^{\left(n\right)}\mathcal A\boldsymbol{z}^{\left(n\right)}\\
0
\end{pmatrix}=\begin{pmatrix}M^{\left(n\right)}\\
\boldsymbol{z}^{\left(n\right)}
\end{pmatrix}\boldsymbol{\zeta}^{\left(n+1\right)}\label{eqIterationFull}
\end{equation}
Here the correction is forced to be perpendicular to the current vector $\boldsymbol{z}^{(n)}$, owing to the rescaling property of homogeneous solutions.  $M^{(n)}_s=2\boldsymbol{z}^{(n)}\mathcal A_s $ is similarly the linear term around $\boldsymbol{z}^{(n)}$, and a correction term $\boldsymbol{\zeta}^{(n+1)}$ is obtained through the pseudo-inverse of $\left(M^{(n)};\boldsymbol{z}^{(n)}\right)$. The vector $\boldsymbol{z}^{(n)}$ is updated according to:
\begin{equation}
    \boldsymbol{z}^{(n+1)} = \frac{1}{\left|\boldsymbol{z}^{(n)}-\lambda_{min}^{(n+1)}\boldsymbol{\zeta}^{(n+1)}\right|}\left(\boldsymbol{z}^{(n)}-\lambda_{min}^{(n+1)}\boldsymbol{\zeta}^{(n+1)}\right)
\end{equation}
With $\lambda_{min}^{(n+1)}$ being the value that minimize the normalized error term $g(\lambda)$,
\begin{align}
\begin{split}
     g(\lambda) =&  \frac{1}{\left|\left|\boldsymbol{z}^{(n)}-\lambda\boldsymbol{\zeta}^{(n+1)}\right|\right|_2} \\&\times\sqrt{\sum_s\left|\left(\boldsymbol{z}^{(n)}-\lambda\boldsymbol{\zeta}^{(n+1)}\right)\mathcal A_s \left(\boldsymbol{z}^{(n)}-\lambda\boldsymbol{\zeta}^{(n+1)}\right)\right|^2} 
\end{split}
\end{align}
A ZPS is adopted from $\boldsymbol{z}^{(n)}$ once it satisfies the constraint to an acceptable level,
\begin{equation}
    \sqrt{\sum_s\boldsymbol{z}^{(n)}\mathcal A_s\boldsymbol{z}^{(n)}} <\epsilon_{\text{ZPS}}\label{eqErrorZPS},
\end{equation} where $\epsilon_{\text{ZPS}}$ is typically close to machine accuracy. An emerging observation is that, for a given vector $\boldsymbol{Z}$, a sequence of subsequent solutions $\boldsymbol{Z}_1,\boldsymbol{Z}_2,...$ can be systematically constructed by repeatedly applying the steps described in Eq.~\eqref{eqIterationFull} onward (even after satisfying Eq.
\eqref{eqErrorZPS}), since each iteration effectively searches for ZPSs in directions orthogonal to those previously identified.\\
Finally, a computational shortcut is available for aggregation of ZPS through the exploitation of a global drive. Since the zero phase gate respects the mirror symmetries between ions along a chain, it can be obtained for a chain where all ions are illuminated homogeneously, as described in Ref. \cite{shapira2020theory}. A global drive represents a version of the problem with a factor $N$ less constraints (and individual degrees of freedom), providing a quadratic speedup with respect to an individual-addressing gate problem.

\section{Algorithm Benchmark}\label{sec: benchmark}
\begin{figure}
\centering{}
\includegraphics[width=1.0\columnwidth]{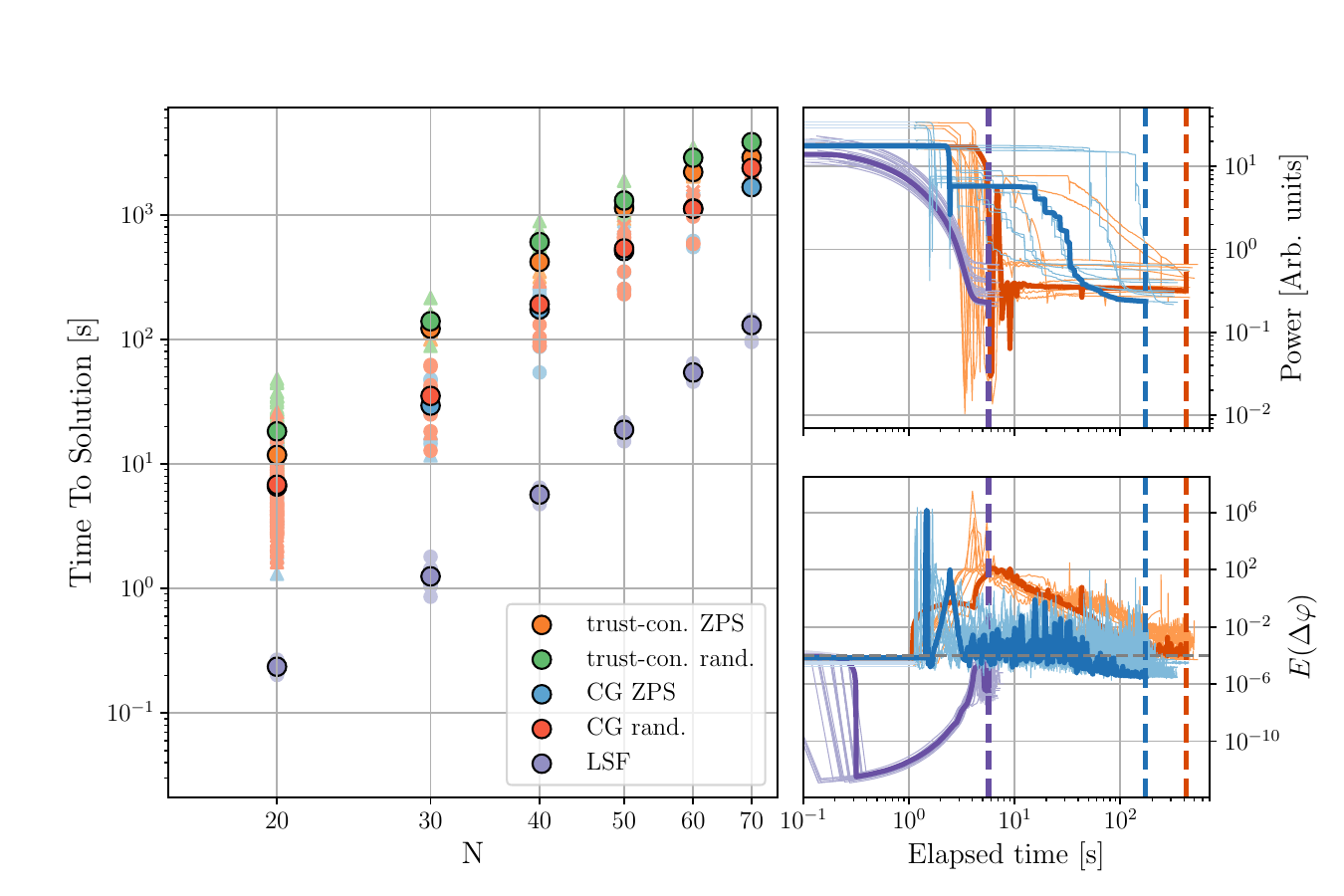} 
\caption{Benchmark comparison of time-to-solution for different optimization algorithms: the LSF method (purple), conjugate-gradient and trust-constraint optimizers starting from a random point (red and green respectively); and same optimizers initialized from $\boldsymbol{R}_\text{conv}$ (blue and orange). Results are shown for ion-crystals of different lengths. Solid markers denote the average run time of gate-design problems, while lighter markers of the same color correspond to individual instances with varying gate times.  
The two right-hand panels show representative optimization trajectories for a 30-ion system: solution power (top) and quadratic error (bottom). These traces compare the LSF method (purple) with trust-constraint (orange) and conjugate-gradient (blue) optimizations initialized at $\boldsymbol{R}_\text{conv}$ starting point for trust-constraint (orange) and conjugate-gradient (blue). Bold curves indicate the averaged trajectory constructed from all recorded optimization runs. Dashed vertical lines mark the average time-to-solution for each algorithm.
}
\label{figBenchmark}
\end{figure}
To evaluate the resource efficiency of our proposed method in generating solutions, we conducted a benchmark comparison between our algorithm and commonly used optimization routines (available in Python's SciPy package ~\cite{virtanen2020scipy}). For each quadratic problem defined by a unique set $\mathcal{A}_s$, we applied the competing algorithms to obtain an optimized solution for Eq.~\eqref{eqOptR}, while maintaining consistent memory and CPU constraints. As alternative optimization approaches, we considered two competing algorithms. The first is the conjugate gradient algorithm, an unconstrained optimization method that minimizes a cost function balancing the phase error
 $\mathcal{E}_\varphi=\sum_{nm} \Delta\varphi_{nm}^2$ , with the drive power term $\mathcal{C}=\lambda\mathcal{E}_\varphi+|\mathbf{r}|^2$. The weighting parameter $\lambda$ is determined based on the predicted optimal drive power, as discussed in the following section. The second is the trust-region constrained optimization method, which enforces the constraint $\mathcal{E}_{\varphi}<\epsilon$ while minimizing $\vb{r}^2$. We provided routines for exact gradient and Hessian calculations wherever they improved the convergence speed of either method. These two specific algorithms outperformed other minimization algorithms in their class on a test problem, and therefore were singled out in a preliminary comparison. 

The set of benchmark problems was created as a two dimensional grid of system size $N$ and gate-time factor $r=T/T_{min}$, with $T_{min}$ the minimal gate-time for a given crystal length (as detailed in section~\ref{sec: physical}). For each parameter pair $P_i={N_i,r_i\cdot T_{min}(N_i)}$, a corresponding set of quadratic matrices ${\mathcal{A}s}^{(i)}$ was derived, defining the constraints for the competing algorithms degrees of freedom. To maintain consistency, the LSF method was initialized at $\boldsymbol{R}_\text{conv}$, a transformed ZPS solution. To ensure fair evaluation, we allowed each competing algorithm to start from two initial conditions: a randomly generated guess and the same $\boldsymbol{R}_\text{conv}$ used in LSF. We tracked execution time, gate fidelity and solution power at each iteration throughout the optimization process until convergence, and compared these performance trends for comparison of the different methods.
Figure~\ref{figBenchmark} summarizes the benchmark results, illustrating the average time to solution $\tau$ versus system size $N$. The averages were computed across varying gate-time ratio parameters $r$, with error bars representing one-sided standard deviations. All algorithms demonstrate a power-law relationship with system size $N$, with the LSF method (purple) showing consistently shorter solution times compared to trust-region constrained optimization and conjugate gradient optimization initiated with either a random guess or $\boldsymbol{R}_\text{conv}$ . The inset provides convergence trends observed for a 30-ion system with different gate durations. Darker lines represent averages of multiple (lighter-colored) instances. Convergence time is defined as the earliest point at which the optimization achieves a value within 5\% of the final result while maintaining an error below the threshold $\epsilon=10^{-4}$. We attribute the improved performance primarily to the explicit control of error through the regulation of $\delta_{\parallel}$ and $\delta_{\perp}$. This is reflected in Fig.~\ref{figBenchmark} as a smooth progression toward the solution, in contrast to the corrective or destructive updates and plateau-like behavior that lead to erratic convergence in gradient-based and trust-region methods.

\section{Summary}

Here we present a systematic framework for the design of large-scale multiqubit entangling gates in trapped-ion systems. We introduce a collection of theoretical and numerical tools, collectively referred to as the \textit{large-scale fast} (LSF) method, which enables the efficient construction of general multiqubit gates in ion crystals comprising up to hundreds of qubits. Building on this framework, we use LSF to investigate the physical properties and scaling behavior of multiqubit gates as the system size increases. In particular, we derive an estimate for the solution norm and, consequently, for the required drive amplitude $\Omega_{\text{nuc}}$, which is shown to depend on the nuclear norm of the target entanglement map $\|\varphi_{nn'}\|_{\text{nuc}}$. Furthermore, we identify a minimal gate time $T_{min}(N)$ that scales linearly with the number of ions in a crystal, below which no high-fidelity multiqubit gate solutions can be found within our framework. \\
Exploiting this method, we conduct a noise-induced gate-error study focusing on three dominant error mechanisms: motional-mode frequency drifts, drive-amplitude jitter, and motional heating. We derive data-driven models for the expected gate error and analyze how these errors scale with increasing chain lengths and participating ions. These models provide a practical link between empirically measured noise levels and the maximum system size that can support gates of acceptable fidelity. Other error sources, particularly those introducing terms that do not commute with the Lamb-Dicke Hamiltonian, require independent treatment via full Hamiltonian simulations. We note that for high-amplitude solutions produced by our method, interactions with the qubit carrier, and consequently higher-order terms in the Magnus expansion, cannot be neglected and must be verified or compensated to ensure valid gate solutions. Such dedicated noise analyses are also essential for assessing the suitability of multiqubit gates in the context of quantum error-correction protocols, especially with regard to noise propagation through large entangling operations and the resulting hierarchy of effective noise operators. \\
Finally, we demonstrate that the proposed optimization routine improves the time-to-solution by approximately an order of magnitude compared to the best available alternatives, rendering the offline compilation of circuits involving tens of ions and many multiqubit gates a realistic task. While further speedups are possible through implementation in compiled languages or GPU-accelerated routines, such optimizations lie beyond the scope of this work. The optimizer itself is robust and readily generalizes to incorporate additional linear and quadratic constraints. In particular, these extensions can be used to relax the current treatment of linear constraints (via null-space transformations) by instead allowing their controlled minimization to an acceptable tolerance, as well as to include additional quadratic error terms in the objective function alongside the quadratic phase constraints~\cite{kirchhoff2024correction,kang2023designing}. Moreover, the framework extends to scenarios with degenerate or overlapping drives, where control fields are applied to multiple ions simultaneously, whether by design or due to crosstalk. 

\begin{acknowledgments}
This work was supported by the Israel Science Foundation and the Israel Science Foundation Quantum Science and Technology (Grants 2074/19, 1376/19 and 3457/21).
\end{acknowledgments}

\bibliography{references}

\appendix

\section{Model derivation} \label{sec: app model}
We detail the derivation of the model used to formulate the optimization problem in Eq. \eqref{eqOptR}. In our analysis below we rely on and generalize the derivations provided in \cite{shapira2020theory}.

We consider a general spectrum of tones which drive independently $N$ trapped ions. 
Without loss of generality, the driving applied to different ions may be regarded as composed  of the same tones, differing only by the tone amplitudes. These tones are placed symmetrically around the single qubit transition frequency, $\omega_0$. The field driving the $n$th ion is,
\begin{equation}
    \begin{split}
    w_n\leri{t}=&\cos\leri{\omega_0 t+\phi_0} \\
    \cdot&\sum_{m=1}^M\left[r_{nm}^{\leri{c}}\cos\leri{\omega_m t} +r_{nm}^{\leri{s}} \sin\leri{\omega_m t}\right].
    \end{split}
\end{equation}
That is, each spectrum contains $2M$ components at frequencies $\left\{\omega_0\pm\omega_m\right\}_{m=1}^M$. The amplitude of the cosine (sine) $m$th tone pair illuminating the $n$th ions is $r_{nm}^{\leri{c}}$ ($r_{nm}^{\leri{s}}$), such that all $N\times M$ tone pairs have the same average phase, which generates a correlated rotation around the Pauli $\sin\leri{\phi_0}\sigma_x+\cos\leri{\phi_0}\sigma_y$ axis. For simplicity we assume that $\phi_0=\pi/2$ such that the relevant Pauli operator is $\sigma_x$. The motional mode phase space trajectories generated by the cosine and sine quadrature are relatively rotated by $\pi/2$.

All in all this is captured by the lab-frame Hamiltonian ($\hbar=1$),
\begin{equation}
    \begin{split}
        H&=H_0+V\\
        H_0&=\sum_{j=1}^N \left[\nu_j \leri{a^\dagger_j a_j+\frac{1}{2}}+\frac{\omega_0}{2}\sigma_z^{\leri{j}}\right]\\
        V&=2\Omega\sum_{n=1}^N \sigma_x^{\leri{n}} \cos\leri{k x_n-\omega_0 t} \cdot\\
        &\sum_{m=1}^M\left[\leri{\boldsymbol{r}_{n}^{\leri{c}}}_m \cos\leri{\omega_m t}+\leri{\boldsymbol{r}_{n}^{\leri{s}}}_m \sin\leri{\omega_m t}\right],
    \end{split}\label{eqHamiltonianLab}
\end{equation}
with $a_j$ the annihilation operator associated with $j$th normal mode of motion, at frequency $\nu_j$, $\sigma_i^{\leri{n}}$ the $i$-Pauli operator acting on the $n$th ion, $k$ the driving field's wave number, $x_n$ the position operator of the $n$th ion and $\Omega$ a characteristic Rabi frequency. Clearly the last parenthesis in Eq. \eqref{eqHamiltonianLab} can be written as a single cosine term with a phase, however this form preserves a crucial aspect of our formulations, i.e. the exclusive linear and quadratic dependence on the amplitudes. Furthermore, here we have assumed that the drive couples to motional modes along one principle direction of the trap, such that the summation on modes is up to $N$ (and not $3N$), this assumption can be easily relaxed \cite{zhu2023pairwise}.

By using a conventional set of approximations, namely the rotating wave approximation in $\Omega/\omega_0$, the Lamb-Dicke approximation, and by neglecting carrier-coupling terms the lab-frame Hamiltonian is converted to the interaction Hamiltonian,
\begin{equation}
    \begin{split}
        H_\text{I}&=\Omega\sum_{n=1}^N \sigma_x^{\leri{n}} \sum_{j=1}^N \eta_j O_j^{\leri{n}} \leri{a_j^\dagger e^{i\nu_j t}+\text{H.c}}\cdot\\
        &\sum_{m=1}^M\leri{r_{nm}^{\leri{c}} \cos\leri{\omega_m t}+r_{nm}^{\leri{s}} \sin\leri{\omega_m t}},
    \end{split}\label{eqHamiltonianInteraction}
\end{equation}
with $O_j^{\leri{n}}$ the normalized participation of the $n$th ion in the $j$th mode of motion, such that $\sum_{j=1}^N O_n^{\leri{j}}O_m^{\leri{j}}=\delta_{n,m}$, and $\eta_j$ the single-ion Lamb-Dicke parameter associated with the $j$th mode of motion (it is sometimes conventional to define $\eta_j^{\leri{n}}=\eta_j O_j^{\leri{n}}$).

The Hamiltonian in Eq. \eqref{eqHamiltonianInteraction} can be rearranged in the form, $H_\text{I}=\sum_{n=1}^N\sum_{j=1}^N \zeta_j^{\leri{n}}\leri{t}\sigma_x^{\leri{n}}a_j+\text{H.c}$, with $\zeta\leri{t}$ a time dependent function read-off directly from Eq. \eqref{eqHamiltonianInteraction}. That is, it has only $\sigma_x$ spin operators and is linear in the mode raising and lowering operators, it is therefore analytically solvable. Specifically its Magnus expansion vanishes after the second order. The resulting unitary evolution operator is the well-known combination of spin-dependent displacement of the motional modes and spin-exclusive correlated rotation,
\begin{equation}
    U=\prod_{j=1}^N D_j\leri{\sum_{n=1}^N\sigma_x^{\leri{n}}\alpha_j^{\leri{n}}}\prod_{n,m=1}^N e^{i\varphi_{n,m} \sigma_x^{\leri{n}}\sigma_x^{\leri{m}} },
    \label{eqUnitary}
\end{equation}
with $D_j\leri{\alpha}=\exp\leri{\alpha a_j^\dagger-\alpha^\ast a_j}$ a displacement operator which translates the $j$th mode by $\alpha_j^{\leri{n}}$, with,
\begin{equation}
    \begin{split}
        \alpha_j^{\leri{n}}&= -i\eta_{j}O_{j}^{\left(n\right)}\Omega \\
        \cdot&\sum_{m=1}^{M}\int _{0}^{t}dt^\prime e^{i\nu_{j}t^\prime}\left[r_{nm}^{\leri{c}}\cos\left(\omega_{m}t^\prime\right)-r_{nm}^{\leri{s}}\sin\left(\omega_{m}t^\prime\right)\right]
    \end{split}
    \label{eqUnitaryDisplacement}
\end{equation}
and entanglement phases,
\begin{equation}
    \begin{split}        
        \varphi_{n,m}&=\boldsymbol{r}_{n}^{T}A_{n,m}\boldsymbol{r}_{m}\\        
        A_{n,m}&=-\Omega^2\sum_{j=1}^{N}\eta_{j}^{2}O_{j}^{\left(n\right)}O_{j}^{\left(m\right)}\begin{pmatrix}A_{j}^{\cos,\cos} & A_{j}^{\cos,\sin}\\
        A_{j}^{\sin,\cos} & A_{j}^{\sin,\sin}
        \end{pmatrix} \\
        \left(A_{j}^{f,g}\right)_{m,l}&=-\int_{0}^{t}dt_{1}\int_{0}^{t_{1}}dt_{2}\sin\leri{v_{j}\left[t_{1}-t_{2}\right]}\\   \cdot&\left[f\left(\omega_{m}t_{1}\right)g\left(\omega_{l}t_{2}\right)+f\left(\omega_{m}t_{2}\right)g\left(\omega_{l}t_{1}\right)\right]
    \end{split}
    \label{eqUnitaryEntanglement}
\end{equation}
with $\boldsymbol{r}_n=\leri{\boldsymbol{r}_n^{\leri{c}}, \boldsymbol{r}_n^{\leri{s}}}$. 

Using the definitions in Eqs. \eqref{eqUnitaryDisplacement} and \eqref{eqUnitaryEntanglement} we observe that mode displacement is linear in the field amplitudes while the two-qubit rotation phase is quadratic in them. While the linear constraints resulting from the former are discussed in Appendix~\ref{sec: app linear}, the quadratic constraints resulting from the latter lead to the optimization problem in Eq. \eqref{eqOptR}.

\section{Gate harmonics and linear constraints}\label{sec: app linear}

We discuss convenient choices for the tones $\left\{\omega_m\right\}_{m=1}^M$ and show how to write the degrees of freedom in a from that by-construction satisfies the linear constraints.

Driving the entanglement operation with a multi-tone representation of the field carries a lot of physical intuition. Specifically, it is beneficial to choose the tones $\omega_m$ in the vicinity of the mode frequency band $\left\{\nu_j\right\}_{j=1}^N$, since the coupling between tones and modes scales inversely with their frequency difference. Thus in our demonstrations above we choose $\omega_1$ slightly below the smallest tone frequency and $\omega_M$ (assuming they are ordered) slightly above the largest frequency, defined precisely below.

We expect the field amplitude to vanish before $t=0$ and after $t=T_\text{g}$, therefore a convenient choice of tones is the harmonic basis, i.e. $\omega_m=\frac{2\pi}{T}h_m$, with $h_m\in\mathbb{N}$ the tone number. We note that the harmonic choice simplifies the evaluation of the integrals in Eqs. \eqref{eqUnitaryDisplacement} and \eqref{eqUnitaryEntanglement}. 

In the main text we state that the optimization problem needs to satisfy both linear and quadratic constraints, however the former may be solved by-construction. Indeed, in order to ensure that at the gate time, $t=T_\text{g}$, the state of the motional mode is the same as in $t=0$ we therefore require that the displacement operatorss in Eq. \eqref{eqUnitary} are unit operators. This ensures that an initial state in which the qubit and motion degrees of freedom are decoupled, remain decoupled after the operation. This is satisfied by requiring that,
\begin{equation}
    \alpha_j^{\leri{n}}\leri{T}=0\text{ }\quad\forall j,n=1,...,N.\label{eqLinearConstraints}
\end{equation}

While Eq. \eqref{eqLinearConstraints} naively contains $N^2$ constraints, they can all be solved by restricting $\boldsymbol{r}_n\in\ker\leri{L}$ for all $n=1,..,N$, with, 
\begin{equation}
    L_{j,m}=\begin{cases}
\int\limits _{0}^{t}dt\cos\left(\nu_{j}t\right)\cos\left(\omega_{m}t\right) & m=1,...,M\\
\int\limits _{0}^{t}dt\sin\left(\nu_{j}t\right)\sin\left(\omega_{m}t\right) & m=M+1,...,2M
\end{cases},
\end{equation}
and $j=1,...,N$. This restriction removes the linear requirement from the optimization problem in Eq. \eqref{eq 6: problem in r}. Furthermore, the matrices $A_{n,m}$ and $A_j$ can be easily transformed to the kernel space of $L$ such that their dimension is reduced and their evaluation is faster. The kernel space can be either found exactly or approximately under some infidelity tolerance \cite{blumel2021efficient}. 

The entanglement operation can be endowed with additional properties that ensure its robustness to various sources of errors and noise such as unwanted coupling to the carrier, and other transitions, pulse timing errors, phonon-mode heating, phonon frequency drifts etc. These may be added as additional rows of the matrix $L$ \cite{shapira2020theory}. 

Out of these properties, decoupling of the carrier transition is useful in order to improve the approximation used to derive Eq. \eqref{eqHamiltonianInteraction}. A convenient way of doing so is by letting the driving fields rise and fall continuously at $t=0$ and $t=T_\text{g}$, respectively, with the choice  $\boldsymbol{r}_n^{\leri{c}}=0$ for all $n=1,..,N$, which further simplifies all of the expressions above. Here we set $\boldsymbol{r}_n\mapsto\boldsymbol{r}_n^{\leri{c}}=0$ and for simplicity identify $\boldsymbol{r}_n$ with $\boldsymbol{r}_n^{\leri{s}}$ and $A_j$ with $A_j^{\sin,\sin}$. We remark that with this choice the drive profile is not time-symmetric, while there are known advantages for using a time-symmetric drive \cite{leung2018robust}.\\
Through the study we apply commonly two additional linear constraints, improving the pulse robustness to timing error ($T_g\to T_g+\delta T$ ) and normal mode drift $\nu_j\to \nu_j +\delta\nu_j$, which we name \textit{Cardioid} and \textit{Nuoid} constraints, respectively, following Ref. \cite{shapira2018robust}. These are derived by enforcing additional constraints on $\alpha^{(n)}_j(T)$. \\
Cardioid constraints are obtained by requiring $\left.\frac{d\alpha_j(T_\text{g})}{d\delta T}=0\right|_{\delta T=0},\forall j$, leading to $2N$ additional linear constraints for the real and imaginary part of each mode:
\begin{align}
\begin{split}
    \Re\left [L^{(\text{car})}_{j,m}\right] = &g(\omega_mT)\cos(\nu_jT)\\
    \Im\left [L^{(\text{car})}_{j,m}\right] =& g(\omega_mT)\sin(\nu_jT)\\
    g(\cdot)=&\left\{\begin{array}{lll}\cos(\cdot),& m=1,...,M\\\sin(\cdot),& m=M+1,...,2M
    \end{array}\right..
\end{split}
\end{align}
Normal mode robustness is obtained by a complementary requirement  $\int_0^T\alpha_j(\tau)d\tau=0$. These results in the following matrix element:
\begin{align}
\begin{split}
    \Re\left [L^{(\text{nu})}_{j,m}\right] = &\frac{\omega_mT}{\omega^2_m-\nu_j^2}-\frac{\left(\omega^2_m+\nu_j^2\right)\cos(\nu_j T)f(\omega_mT) }{\left(\omega^2_m-\nu_j^2\right)^2}\\&+\frac{ 2\nu_j\omega_m\sin(\nu_j T)g(\omega_mT)}{\left(\omega^2_m-\nu_j^2\right)^2}\\
    \Im\left [L^{(\text{car})}_{j,m}\right] =&\frac{\omega_mT}{\omega^2_m-\nu_j^2}+\frac{2\nu_j\omega_m\cos(\nu_j T)g(\omega_mT) }{\left(\omega^2_m-\nu_j^2\right)^2}\\&+\frac{\left(\omega^2_m+\nu_j^2\right)\sin(\nu_j T)f(\omega_mT)-2\nu_j\omega_m}{\left(\omega^2_m-\nu_j^2\right)^2}\\\\
    f(\cdot),g(\cdot)=&\left\{\begin{array}{lll}\sin(\cdot),\cos(\cdot),& m=1,...,M\\\sin(\cdot),\sin(\cdot),& m=M+1,...,2M
    \end{array}\right.
\end{split}
\end{align}

\section{Trivial solution in the adiabatic limit}\label{sec: app adiabatic}
In the slow gate limit, $\Delta\bar{\nu} T_\text{g}\gg1$ satisfying the quadratic constraints is trivial and any bipartite qubit-qubit coupling can be achieved. Here we prove this directly by constructing such a solution. In this limit it is helpful to choose spectral tones containing frequencies of the form $\omega_s\mapsto\omega_{j,s}=\nu_j+\frac{2\pi}{T}s$ with $s\in\mathbb{N}$. We have doubled the index $s$ to $\leri{j,s}$ for convenience. Since $T_\text{g}$ is large we can safely assume that the tone $\omega_{j,s}$ couples exclusively to mode $j$ and satisfies the linear constraints by construction. Furthermore, with this choice the matrices $A_j$ become diagonal, i.e each of the tones coupled to mode $j$ contributes to the entanglement phase independently, and scales as $s^{-1}$. In other words $\leri{A_j}_{\leri{j^\prime,s^\prime},\leri{j^{\prime\prime},s^{\prime\prime}}}\propto\delta_{j,j^\prime}\delta_{j,j^{\prime\prime}}\delta_{s^\prime,s^{\prime\prime}}\leri{s^\prime}^{-1}$ \cite{shapira2018robust}.

We are required to generate $N\leri{N-1}/2$ bipartite entanglement phases $\varphi_{n,m}$, with $1\leq n<m\leq N$. We do so by designating a unique tone to each phase. Since these tones do not interfere (whether they are driving distinct or the same mode of motion) then we simply need to scale the drive amplitudes accordingly. Specifically, to satisfy the constraint on $\varphi_{n,m}$ we choose an arbitrary mode which couples to both ions. Without loss of generality this can always be the center of mass mode, $j_\text{COM}$. We drive both ions with the same tone $\omega_{j,s}$ setting $j=j_\text{COM}$ and $s=N\cdot n+m$, and set its amplitude driving both ions to be $\sqrt{\frac{\leri{N\cdot n+m}\varphi_{n,m}}{O_{j_\text{COM}}^{\leri{n}}O_{j_\text{COM}}^{\leri{m}}}}$. 

\section{Effects of robustness on gate power}\label{sec: app nuc linear}
In the main text we show that the total gate power, given by the nuclear norm $\Omega_\text{nuc}$ depends linearly on the nuclear norm of the target coupling matrix (and other parameters), see Eq. \eqref{eqNuclearNorm}, with $k$ the linear proportionality constant. Here we provide a heuristic argument for the numerical dependence of $k$ on the robustness properties of the entanglement operation, highlighted in the main text.

The framework for analytic analysis is multi-tone adiabatic gate that couple to single modes of motion, as was presented in Ref. \cite{shapira2018robust}. In that work drive tones pairs are symmetrically positioned around the red and blue sideband of the motional mode of choice, i.e. with frequencies, $\omega_{n,\pm}=\omega_0\pm\left(\nu+n\xi\right)$, with $\omega_0$ the qubit transition frequency, $\nu$ the frequency of the mode of motion of choice, $\xi$ an additional detuning such that the gate time is given by $2\pi\xi^{-1}$ an $n\in\mathbb{Z}$. With this formulation, the gate is designed by constraints on the amplitudes driving each tone pair.

\begin{figure}
\centering{}\includegraphics[width=1\columnwidth]{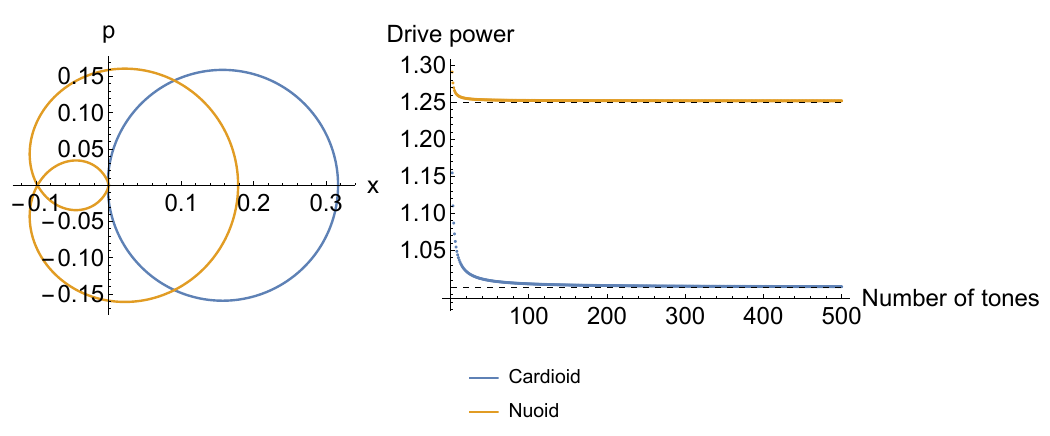} \caption{Left: Phase space trajectory of two-qubit robust gates in the limiting case of infinite tones. At this limit the Cardioid gate (blue) recovers the circular trajectory of the canonical MS gate, which explains the vanishing cost of its robustness, while the Nuoid gate (orange) has a distinct shape. Right: Drive power of the Cardioid (blue) and Nuoid (orange) gates as a function of number of tones. Indeed the Cardioid gate power converges to 1, while the Nuoid gate power converges to 1.25, i.e. a finite additional cost of the robustness.}
\label{figLagrangian}
\end{figure}

Specifically, gates that are robust to timing errors (among other sources of error) are generated by satisfying, $\sum_n r_n^2/ n=1$ and $\sum_n r_n=0$ with the former ensuring a desired entangling phase and the latter ensuring the robustness property (coined \textit{Cardioid} in Ref. \cite{shapira2018robust}). The total power, $\sum_n r_n^2$ is to be minimized subject to these constraints. We do so in a Lagrange multiplier framework, by minimizing the Lagrangian,
\begin{equation}
    L_\text{Car}=\sum_n r_n^2-\lambda\leri{\sum_n\frac{r_n^2}{n}-1}-\mu\leri{\sum_n r_n}.\label{eqLagCar}
\end{equation}
One then solves the equations $\partial_{r_n}L_\text{Car}=0$ under these constraints. At the limit of the number of tones approaching infinity, the solution to these equations converges to a power $\sum_n r_n^2=1$, which is the same result obtained in absence of the robustness constraint. That is, presumably this constraint does not incur a overhead of gate drive power. However, one notices that this is true only in the unphysical limit of infinite driving tones and that this limiting solution in fact recreates the standard non-robust version of the MS gate, shown in Fig. \ref{figLagrangian}.
A similar example is presented for gates that are robust to calibration errors of the motional mode frequency, given by the constraints, $\sum_n r_n^2/n=1$ and $\sum_n{r_n/n=0}$ (coined \textit{Nuoid} in Ref. \cite{shapira2018robust}). Similarly, a Lagrangian is formed,
\begin{equation}
    L_\text{Nu}=\sum_n r_n^2-\lambda\leri{\sum_n\frac{r_n^2}{n}-1}-\mu\leri{\sum_n \frac{r_n}{n}},\label{eqLagNu}
\end{equation}
and solution to $\partial_{r_n}L_\text{Nu}=0$ are found under these constraints. Here the we get, at an infinite number of tones, a drive power of $\sum_n r_n^2\approx1.25$, showing an actual overhead in terms of drive power in order to implement this robustness property, shown as well in Fig. \ref{figLagrangian}.

\section{Von-Neumann entropy calculations}\label{sec: app von}
There are many measures that can be used to quantify the entanglement generated by $U_\text{MQ}$. Here, for simplicity, we choose the Von-Neumann single-qubit entropy, i.e. we compute $\overline{S}_\text{VN}\leri{\varphi}=\sum_n S_{\text{VN},n}=-\sum_n \rho_n\log_2\leri{\rho_n}$, where $\rho_n$ is a density matrix formed by tracing out the $n$th qubit. We perform this computation over the state $U_\text{MQ}\ket{+}^{\otimes N}$, where $\ket{+}$ is the $+1$ eigenvalue of $X_n$.

Considering $\rho_n$, we have, 
\begin{equation}
\begin{split}
    \rho_n&=\frac{1}{2}\exp\leri{i\sum_{j,k\neq n}\varphi_{j,k}Z_jZ_k}\cdot\\
    &\big(e^{i\sum_{k\neq n}Z_k}\ketbra{+_{~n}}{+_{~n}}e^{-i\sum_{k\neq n}Z_k}+\\
    &e^{-i\sum_{k\neq n}Z_k}\ketbra{+_{~n}}{+_{~n}}e^{+i\sum_{k\neq n}Z_k}\big)\\
    &\exp\leri{-i\sum_{j,k\neq n}\varphi_{j,k}Z_jZ_k}
\end{split}
\end{equation}

Where $\ket{+_{~n}}$ is the original product state with the $n$th qubit absent. and we decomposed the state of the $n$th qubit to $\ket{+}=\leri{\ket{0}+\ket{1}}/\sqrt{2}$, and then replaced the values of $Z_n$ with $\pm1$ accordingly. The resulting density matrix is an equal weight sum of two non-orthogonal states, thus its non-zero eigenvalues are given by $\lambda_\pm=(1\pm|S|)/2$ with $S$ the overlap between the states in absolute value. It is trivial to show that here $S=\left|\prod_{k\neq n}\cos^2\leri{2\varphi_{n,k}}\right|\approx1-2\sum_{k\neq n}\varphi_{n,k}^2$, where the approximation is a leading order expansion in small values of $\varphi$.

Thus the two eigenvalues are $\sum_{k\neq n}\varphi_{n,k}^2$ and $1-\sum_{k\neq n}\varphi_{n,k}^2$

We get,
\begin{equation}
    S_{\text{VN},n}=-\rho_n\log_2\rho_n\approx\sum_{k\neq n}\varphi_{n,k}^2,
\end{equation}
where again the approximation is in leading order in $\varphi$. Thus we obtain, $\overline{S}_\text{VN}\leri{\varphi}=\sum_{n,k}\varphi_{n,k}^2$, which is equal to the Frobenius norm squared of the matrix $\varphi$.

\section{Choice of effective degrees of freedom}\label{sec: DOFs}
\begin{figure}
\centering{}\includegraphics[width=0.9\columnwidth]{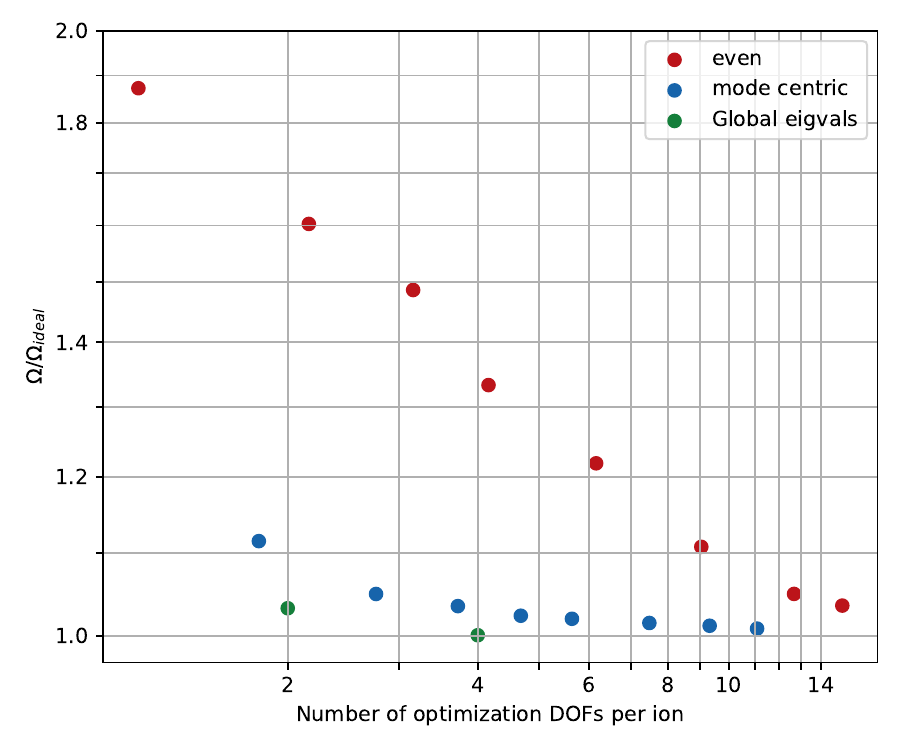} \caption{Comparison of obtained solution amplitude $\Omega$ for drives where the number of optimization variables was reduced from $M\sim  N^2$ to $2k N$, along different methods - even dilution (red), mode-centric dilution (blue) and mode-phase eigenvalues (green). at $k=2$ optimization DOFs, the eigenvalues method is the first to produce a solution of the same $\Omega$ as a system without DOF reduction.}
\label{figDOFsCompare}
\end{figure}
While fast gates are generally desirable, there are scenarios where only slower gates are feasible, particularly when drive power is limited or error-budget constrained.  With the nuclear norm of an entangling map scaling with the number of participating ions, the minimum feasible power-limited gate time scales as $N^2$: one factor of $N$ from $T_{\text{min}}$, and the other from limited optical power, as $|r|\sim N/T_\text{g}$. As the number of Fourier components is linear in $T_g$, a naive gate construction would require $M=\mathcal{O}(N^2)$ frequency components, even though only $\mathcal{O}(N)$ are likely required to control the phase of $N$ modes. This translates to a steep increase in both hardware complexity and LSF memory usage, as each ion-pair constraint matrix has size $M^2$. A cost-efficient heuristic for eliminating less effective degrees-of-freedom a-priory is therefore required.\\
 DOF reduction can be achieved either by zeroing selected drive amplitudes $\Omega_m^{(n)}$ in advance, or by introducing a dimensionality-reducing linear map 
 $L\in\mathcal{R}^{M'\times M}$ with $M'<M$ such that $\boldsymbol{r}_n=LK\boldsymbol{\Omega^{(n)}}$. The first approach compresses the drive signal representation; the second reduces the size of the quadratic matrices while preserving the linear constraint space defined by matrix $K$. Particularly, We considered three heuristics:
 \begin{enumerate}
     \item keeping exactly $2k$ evenly-distributed tones between each two modes (even dilution).
     \item  keeping the $2k$ nearest tones around each mode (mode-centric dilution).
     \item Constructing $L$ with $2kN$ rows based on the eigenvectors corresponding to the $k$ largest positive and $k$ largest negative eigenvalues of each mode-phase matrix $KA_jK^T$ (see Appendix~\ref{sec: app model}, Eq. \eqref{eqUnitaryEntanglement}) (eigenvalues dilution).
\end{enumerate}
To evaluate the effectiveness of DOF reduction, we applied the proposed methods to a 25-ion gate design problem and compared the resulting solutions to a reference solution using the full set of 805 tones, corresponding to a gate time of $T_g=6\cdot\Delta\nu_<\approx 32\bar{\Delta\nu}$. This gate time was chosen to allow for additional robustness constraints, resulting in 703 optimization DOFs within the kernel of the linear constraint matrix - approximately 28 DOFs per mode.
Fig. \ref{figDOFsCompare} shows the resulting drive amplitudes, normalized by the amplitude of the reference (non-diluted) solution $\Omega_{ref}$,  for the three dilution strategies: evenly spaced tones (blue), mode-centric tones (orange), and the eigenvalue-based method (green), as a function of the dilution parameter $k=\frac{\#DOFs}{N}$. In the even and mode-centric strategies, tone selection is applied prior to imposing linear constraints, resulting in fewer actual optimization DOFs than the nominal $k$ value suggests.
Among the three, the eigenvalue method most rapidly recovers the reference amplitude $\Omega_{ref}$,using the two largest positive and two largest negative eigenvectors per mode. This highlights the benefit of selecting the most effective DOFs for independently steering each mode’s phase contribution to
 $\varphi_{n,m}$. The mode-centric method performs slightly worse, consistent with the fact that the eigenvalue-derived DOFs are concentrated near their respective modes in frequency space. The even dilution method performs significantly worse, demonstrating that enforcing a uniform distribution of tones across the spectrum leads to suboptimal amplitude solutions. Based on these results, we adopt the eigenvalue method as the default tone dilution strategy for large-scale gate optimization problems.

\end{document}